\documentclass[aps,notitlepage,superscriptaddress,floatfix,pra,twocolumn,nofootinbib]{revtex4-1}

\usepackage{lipsum}
\usepackage{graphicx}
\usepackage{amsmath,amssymb}
\usepackage{braket}
\usepackage{mathtools}
\usepackage{hyperref}
\usepackage{multirow}
\usepackage{color}
\usepackage[normalem]{ulem}
\usepackage{amsfonts}
\usepackage{float}
\usepackage{pdfpages} 
\makeatletter
\usepackage{subfigure}
\usepackage{soul, xcolor}
\usepackage{dsfont}
\usepackage{MnSymbol}

\setstcolor{red}

\def\bra#1{\mathinner{\langle{#1}|}}
\def\ket#1{\mathinner{|{#1}\rangle}}

\def\beq{\begin{equation}}
\def\eeq{\end{equation}}
\def\bea{\begin{eqnarray}}
\def\eea{\end{eqnarray}}

\usepackage{pdfpages} 
\makeatletter
\AtBeginDocument{\let\LS@rot\@undefined}
\makeatother

\begin{document}

\title{Mixed-state learnability transitions in monitored noisy quantum dynamics}

 \author{Hansveer Singh}
\affiliation{Max Planck Institute for the Physics of Complex Systems, 01187 Dresden, Germany}


\author{Romain Vasseur}
\affiliation{Department of Theoretical Physics, University of Geneva, 24 quai Ernest-Ansermet, 1211 Geneva, Switzerland}

\author{Andrew C. Potter}

\affiliation{Department of Physics and Astronomy, and Quantum Matter Institute, University of British Columbia, Vancouver, BC, Canada V6T 1Z1}

\author{Sarang Gopalakrishnan}
\affiliation{Department of Electrical and Computer Engineering, Princeton University, Princeton, NJ 08544, USA}

\begin{abstract}

We consider learnability transitions in monitored quantum systems that undergo noisy evolution, subject to a global strong symmetry---i.e., in addition to the measuring apparatus, the system can interact with an unobserved environment, but does not exchange charge with it. As in the pure-state setting, we find two information-theoretic phases---a sharp (fuzzy) phase in which an eavesdropper can rapidly (slowly) learn the symmetry charge. However, because the dynamics is noisy, both phases can be simulated efficiently using tensor networks. Indeed, even when the true dynamics is unitary, introducing noise by hand allows an eavesdropper to efficiently learn the symmetry charge from local measurements, as we demonstrate. We identify the fuzzy phase in this setting as a mixed-state phase that exhibits spontaneous strong-to-weak symmetry breaking.

\end{abstract}
\vspace{1cm}

\maketitle

\section{Introduction}

Monitored quantum dynamical systems---in which unitary evolution is interspersed with measurements---exhibit phase transitions as the rate of measurement is increased. Although the crucial role of measurements in quantum evolution has been appreciated from the advent of quantum mechanics, these measurement-induced phase transitions (MIPTs) were first discovered relatively recently in studies of random quantum circuits~\cite{PhysRevX.9.031009, PhysRevB.98.205136,Potter_2022,annurev:/content/journals/10.1146/annurev-conmatphys-031720-030658}. Since then, they have been studied in a variety of settings~\cite{PhysRevX.10.041020, PhysRevLett.125.030505, PhysRevB.103.104306, PhysRevB.103.174309, PhysRevB.100.134306, PhysRevB.99.224307, PhysRevB.104.104305, PhysRevB.100.064204, PhysRevLett.125.070606, PhysRevB.101.060301, PhysRevX.11.011030, lavasani_measurement-induced_2021, PhysRevResearch.3.023200, PhysRevResearch.2.013022, PhysRevB.102.014315, PhysRevB.102.054302, PhysRevB.104.155111, PhysRevB.103.224210, PhysRevLett.126.060501,PRXQuantum.2.040319, PhysRevB.106.134206, PhysRevLett.126.170503, PhysRevB.106.144313, PhysRevLett.128.010604, PhysRevB.104.094304, PhysRevLett.128.050602, PhysRevB.108.214302, jian2021quantumerroremergentmagnetic, PhysRevResearch.4.023146, PhysRevLett.128.130605, Sierant2022dissipativefloquet, 10.21468/SciPostPhysCore.5.2.023, PhysRevLett.130.220404, PhysRevB.109.125148, PhysRevResearch.6.033220, PRXQuantum.4.030333, PhysRevB.110.064309, PhysRevB.110.064301, PhysRevB.110.L060202, PhysRevB.108.184302, PhysRevB.107.L220204, PhysRevB.108.184204, PhysRevB.109.014303, PhysRevB.107.L201113, PRXQuantum.5.030329, PhysRevB.110.054308, PhysRevResearch.6.023176, PhysRevLett.130.120402, PhysRevLett.132.240402, PhysRevLett.130.230401, PhysRevB.107.L220201, PhysRevB.110.045135, PhysRevB.108.L041103, PhysRevB.107.094309,PhysRevB.110.064323,PhysRevB.107.014308,PhysRevB.107.224303,PhysRevLett.131.060403,PhysRevLett.131.220404,PhysRevB.108.104310,PhysRevB.108.165126,PhysRevB.108.104203,PhysRevB.107.064303, Li_2023,PhysRevB.108.L020306,JianShapourianBauerLudwig2023,
PhysRevX.13.041045,PRXQuantum.4.040332}, including circuits with symmetries~\cite{BAO2021168618,PhysRevX.12.041002,PhysRevB.108.054307,GuoJianFosterLudwigKeldysh2024,MirlinGornyiEtAlKeldysh2024}, free fermionic systems~\cite{10.21468/SciPostPhys.7.2.024,PhysRevResearch.2.023288,PhysRevLett.126.170602,PhysRevX.11.041004,PhysRevX.13.041046}, etc. MIPTs are most directly understood as transitions in the entanglement structure of the state of a quantum system \emph{conditional} on a set of measurement outcomes. Since any entanglement measure of a quantum state is a nonlinear function of its density matrix, detecting an MIPT through entanglement requires one to compute quantities such as the purity of a quantum state $\mathrm{Tr}(\rho_{\mathbf{m}}^2)$ conditional on a particular set of measurement outcomes $\mathbf{m}$. To compute such quantities, one needs to regenerate the same measurement history many times, incurring an exponential post-selection overhead. This post-selection overhead was thought to make MIPTs unobservable in a scalable way. 

\begin{figure}[!b]
    \centering
    \includegraphics[width=0.5\textwidth]{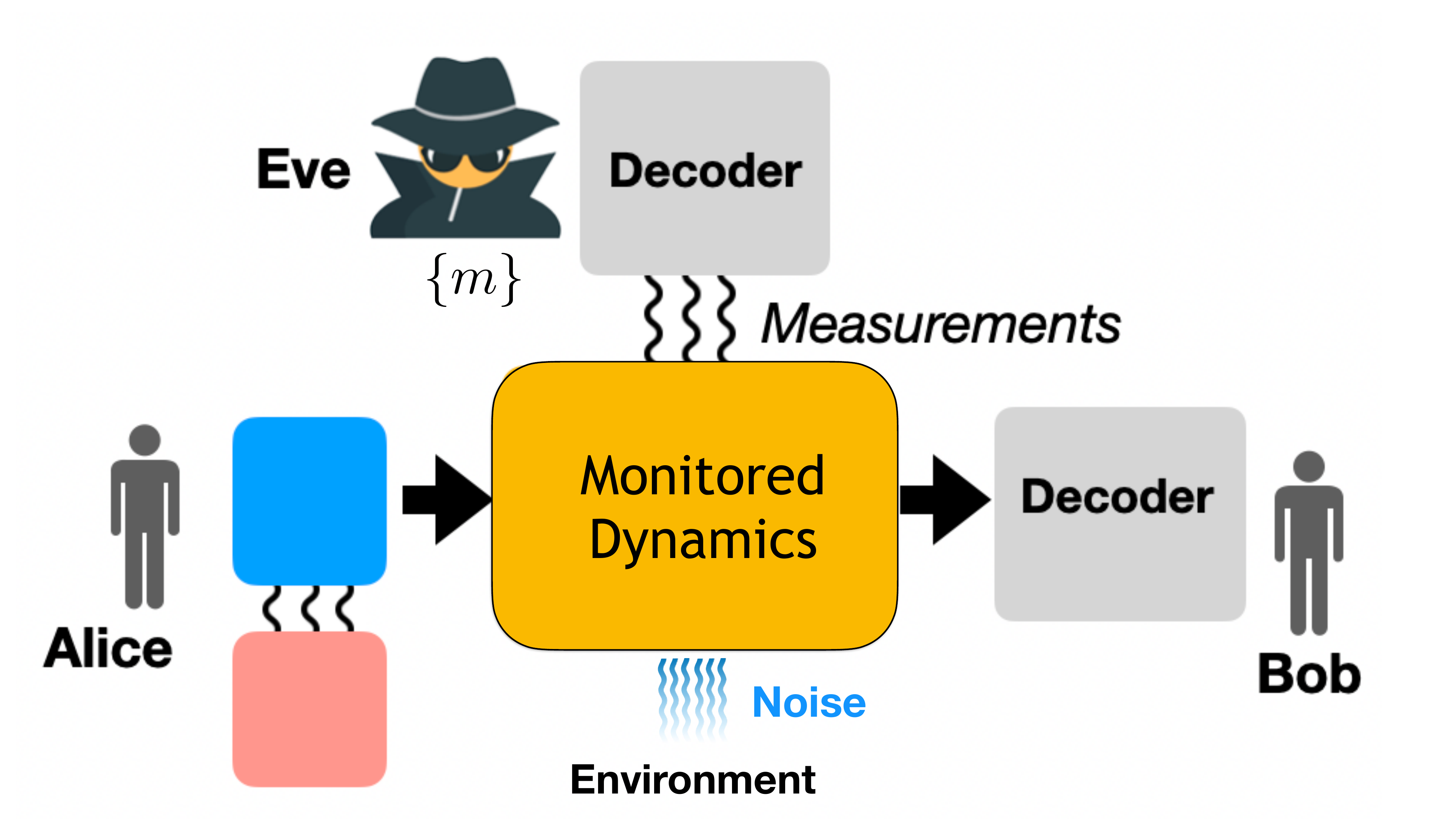}
    \caption{{\bf Setup.} Alice is encoding some quantum information in a superposition of different charges of a global symmetry, and sending it to Bob through a strongly-symmetric monitored noisy quantum evolution. We are interested in the mixed state learnability transitions characterized by how efficiently the observer (Evesdropper) Eve can extract information about the charge of the system.   
     }
    \label{fig:setup}
\end{figure}

In the past two years, various attempts to overcome the post-selection barrier have been proposed and experimentally implemented. The general principle behind these schemes is to use the classical measurement record collected during the experiment to predict (or manipulate) the final state of the system. For example, on the low-measurement side of the standard MIPT in random circuits, the dynamics is an emergent error correcting code~\cite{PhysRevX.10.041020,PhysRevLett.125.030505}. Inside this code space, when it exists, the effect of the hybrid measurement dynamics is purely unitary; to decode, one can implement the inverse unitary. However, \emph{finding} the inverse unitary appears to be a computationally hard task, requiring one to simulate the dynamics (conditioned on the observed measurement record) on a classical computer. Moreover, any imperfections in the measurement record cause decoding to fail. 
Thus, for the standard MIPT, this strategy replaces the post-selection problem with an intractable decoding problem. 

For dynamics with symmetries, the situation is more optimistic.
It has been fruitful to take the perspective of \emph{learning}~\cite{PhysRevLett.129.200602, PRXQuantum.5.020304,agrawal2023observingquantummeasurementcollapse} from the measurement record rather than that of decoding. Consider, for concreteness, a circuit with a conserved scalar charge~\cite{PhysRevX.12.041002}: Alice picks a state of a definite charge $Q$, and initializes the dynamics with that state. Eve performs \emph{local} measurements of the charge on some fraction $p$ of the sites at each time step. Eve's task is to use the measurement outcome data to infer $Q$. This problem has an information-theoretic ``sharpening'' threshold $p_\#$: when $p > p_\#$, Eve can reliably infer $Q$ after measuring for a time $O(\log t)$, but when $p < p_\#$, she needs to measure for a time $O(t)$ to do so. Even when inference is information-theoretically possible, it might be computationally hard: the optimal learning algorithm naively requires Eve to run the dynamics for each candidate value of $Q$ on a classical computer, forcing the observed measurement outcomes, and pick the likeliest $Q$. However, even sub-optimal strategies can have sharp learning thresholds (lower bounded by $p_\#$).   
In particular, the ``stat mech'' decoder in Ref.~\onlinecite{PhysRevLett.129.200602} was shown to succeed with probability 1 above a threshold measurement rate. By contrast, imperfect strategies appear not to have sharp thresholds for the true MIPT.

In this work we point out that the success of the stat mech decoder is a manifestation of a more general feature: the existence of information-theoretic learnability transitions in \emph{noisy} quantum dynamics. Noisy quantum dynamics is (under some assumptions) believed to be easy to simulate~\cite{Noh2020efficientclassical,schuster2024polynomialtimeclassicalalgorithmnoisy}; therefore, noisy information-theoretic transitions are computationally tractable. Indeed, noise can be used as a computational \emph{resource}: even when the eavesdropper has a full description of the unitary evolution operator, ``forgetting'' (or marginalizing over) some of this information---or equivalently, replacing some of the unitary gates with noisy quantum channels---can render the dynamics tractable using matrix-product operator (MPO) methods~\cite{SCHOLLWOCK201196}, while only modestly increasing the learning threshold. This perspective provides an alternative, and much more general, way of understanding the ``stat mech decoder.'' We show that this strategy can be used to study learning transitions for nonabelian symmetries (specifically $U(1)\rtimes \mathbb{Z}_2$). In the former case, we present clear numerical evidence that the dynamics at the charge-learning transition is efficient to simulate. 
We conjecture that (at least for the groups $U(1)$ and $U(1)\rtimes \mathbb{Z}_2$) 
any amount of marginalizing makes the dynamics classically tractable: although the existing proofs that noisy dynamics is classically simulable do not include projective measurements, for these symmetries we find no evidence that measurements increase the simulation complexity. 

Studying learnability transitions in noisy dynamics also allows us to connect these transitions to the phenomenon of spontaneous strong-to-weak symmetry breaking (SW-SSB), which has been extensively explored in recent work~\cite{You24weaksym, Wang24strtowksym}. In order for learnability transitions to be well-defined, we require that the dynamics obey a strong symmetry: i.e., each measurement projector and each Kraus operator of the noise should commute with the global symmetry. In the weak-measurement phase, although there is a global strong symmetry, operators that are charged under this symmetry develop a form of long-range order diagnosed by the fidelity correlator. Physically, the fidelity correlator diagnoses whether the state of the system changes in a detectable way if one moves a unit of charge halfway across the system; in effect, the fidelity correlator vanishes when the measurement outcomes fix the local charge profile (i.e., in the sharp phase), and remains nonzero otherwise. We comment on how this perspective helps us understand the fuzzy phase in dimensions greater than one. 

\section{Charge learning with noise}

We begin by formulating the learnability problem with noise.
This discussion closely follows previous treatments of the charge-sharpening or learnability transition~\cite{PhysRevLett.129.200602, PRXQuantum.5.020304}, just rephrased to make it clear that unitarity is not essential to define this phenomenon.
We will consider a general discrete-time dynamical process consisting of unitary operators, measurements, and noise channels. (The extension to continuous-time dynamics is straightforward.) The dynamics has a global strong symmetry under a group $G$, where each element $g \in G$ acts on the state space through the matrix $M_g$. The symmetry is associated with a set of charges $Q_i$ that label the irreducible representations of $G$. The strong symmetry means that each unitary gate, as well as each Kraus operator for the measurements or the noise, commutes with each $M_g$. Therefore, a mixed state of the system that is symmetric (i.e., $G$-invariant) at some time will remain symmetric at all times; moreover, a density matrix that is initially a definite eigenstate of all the $Q_i$ will remain one under time evolution, and its eigenvalues (charges) will not change. 

\begin{figure*}[!t]
    \centering
    \includegraphics[width=\textwidth]{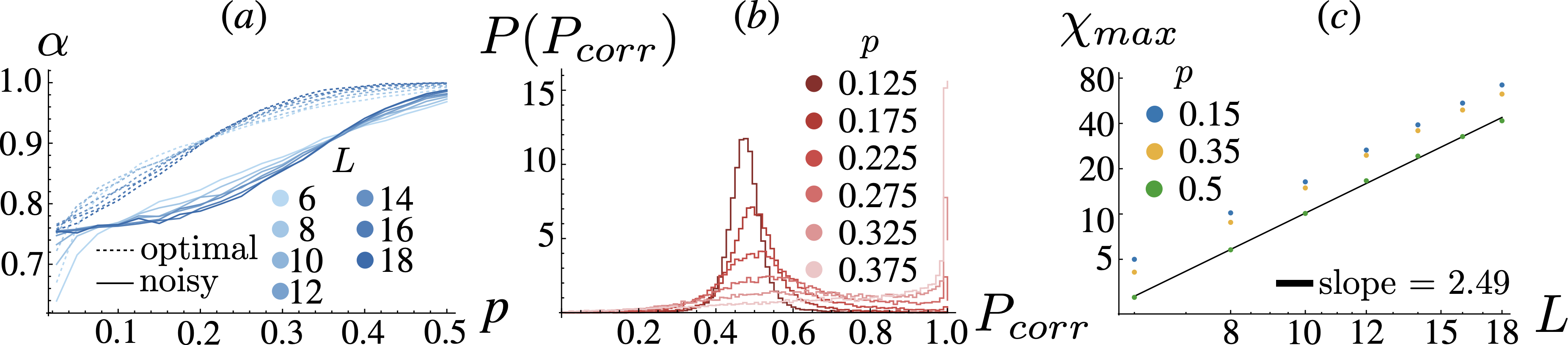}
    \caption{{\bf $\mathbf{U(1) \rtimes Z_2}$ optimal and noisy decoders.} (a) Accuracy, $\alpha$, of the optimal decoder (dashed lines) and noisy decoder (solid lines) for various system sizes. (b) The distribution of the probability of finding the correct charge label, $P_{corr}$ for $L=18$ for various measurement rates obtained with the noisy decoder. (c) System size dependence of the maximum bond dimension reached after $L^2$ time steps of the TEBD simulation for the noisy decoder. The straight line corresponds to the best linear fit. 
     }
    \label{fig:mainfig}
\end{figure*}

The learnability problem can be phrased as follows. Suppose the system is initialized in a state $\sigma^0_Q$ drawn from a set of states each with definite charge $Q$; 
each charge $Q$ occurs with probability $p_Q$. Eve does not know the initial state, and models her classical uncertainty by taking the initial state to be the density matrix $\sigma = \sum\nolimits_Q p_Q \sigma^0_Q$. Conditional on Eve's observing a classical measurement record $\mathbf{m}$, this initial state evolves to $p(\mathbf{m}) \sigma^{\mathbf{m}} = K_{\mathbf{m}} \sigma K^\dagger_{\mathbf{m}}$, where each $K_{\mathbf{m}}$ is a sequence of channels (describing noisy dynamics and/or measurements where the outcome is lost or discarded) and Kraus operators (describing measurements whose outcomes were recorded). The state $\sigma^{\mathbf{m}}$ has unit trace, since the trace of the r.h.s.~is the Born probability $p(\mathbf{m})$ of the measurement record $\mathbf{m}$. Each initial state $\sigma^0_Q$ evolves to the un-normalized state $\tilde\sigma^{\mathbf{m}}_Q$; by the Born rule, $\mathrm{Tr}(\tilde\sigma^{\mathbf{m}}_Q) = p(\mathbf{m}|Q)$, the conditional probability of observing the record $\mathbf{m}$ for an initial state $\sigma^0_Q$. In terms of $\sigma^{\mathbf{m}}_Q \equiv \tilde\sigma^{\mathbf{m}}_Q/p(\mathbf{m}|Q)$, the evolution maps
\beq
\sigma \mapsto \sigma^{\mathbf{m}} = \sum\nolimits_Q \frac{p_Q p(\mathbf{m}|Q)}{p(\mathbf{m})} \sigma^{\mathbf{m}}_Q.
\eeq
The probability factor in front of each state is precisely the conditional probability $p(Q|\mathbf{m})$. Therefore, the charge entropy of the post-measurement state represents the intrinsic uncertainty about $Q$ that remains after Eve has made her measurements. Since measurements reduce uncertainty and the dynamics does not generate any new uncertainty, the distribution $p(Q|\mathbf{m})$ will eventually sharpen to a delta function. 
%
%
In other words, $\sigma_{\mathbf{m}}$ will have definite values of the symmetry charges after characteristic time $t_{\#}$, which will depend on the system size $L$~\footnote{One can imagine cases, such as distinguishing between a charge of $+q$ and $-q$ for measurements of $U(1) \rtimes Z_2$, where both states give precisely the same measurement outcomes, so the system never sharpens. We will not be considering these cases.}. After $t_{\#}$, Eve, in principle, has access to the symmetry charges of the state $\sigma^{\mathbf{m}}$ and can confidently predict the outcomes of future measurements of these global symmetry charges. Learnability transitions correspond to parametric changes in $L$-dependence of the sharpening time $t_{\#}(L)$ as $L \to \infty$. To have a well-defined learnability transition, we will scale $L, t \to \infty$ so that the state $\sigma^{\mathbf{m}}$ is reliably sharp at measurement rates above some critical measurement rate $p_{\#}$, but not otherwise. 

\subsection{Optimal and noisy strategies}

Eve's task is to predict the outcome of a future measurements of global symmetry charges in the state $\sigma^{\mathbf{m}}$ from the measurement record $\mathbf{m}$. An optimal strategy for doing this is to simulate the dynamics, implementing the measurements through the Kraus operators that correspond to the observed outcomes (a task in the complexity class of PostBQP that is generally hard for both classical and quantum computers~\cite{doi:10.1098/rspa.2005.1546}). This procedure reconstructs $\sigma^{\mathbf{m}}$. If $\sigma^{\mathbf{m}}$ has a sharp eigenvalue of the desired charge, Eve predicts that value; otherwise she guesses the likeliest value of the charge from $\sigma^{\mathbf{m}}$. The computational complexity of this procedure depends on how difficult it is to simulate the dynamics on a classical computer. When the dynamics consists of generic unitaries and projectors alone, simulating the dynamics appears to be computationally hard below a critical measurement rate $p_c$, which (in general) lies inside the sharp phase. Thus, the optimal strategy can be computationally hard even when learning is \emph{in principle} possible. On the other hand, if the underlying dynamics is noisy, it is expected (but not proven) that Eve will be able to simulate the dynamics efficiently for all measurement rates~\footnote{Establishing this rigorously requires extending recent results on simulating noisy quantum dynamics to systems with measurements.}. 

Even when the true dynamics is unitary, Eve can artificially \emph{introduce} ignorance to simplify the decoding task. This can be done by throwing out information about some gates or some measurement outcomes. Formally, one does this by replacing the ignored gates and/or measurements with strongly symmetric quantum channels. The resulting conditional state $\sigma^{\mathbf{m}}$ represents the extent of Eve's knowledge about the state based on the information she retained. Given this knowledge, the best Eve can do to infer the charges is to implement this noisy evolution (fixing measurement outcomes) on a classical computer, and follow the strategy outlined above. Because accessible information is monotonic under quantum channels, this ``noisy decoder'' is less correlated with the ground truth values of charge than the truly optimal decoder. At least for sufficiently large values of noise, however, it will be computationally efficient. 

\subsection{Example: $U(1)\rtimes \mathbb{Z}_2$}

In the simplest case (that of a scalar $U(1)$ charge) the noisy strategy we outlined above was already implemented as the ``stat mech'' decoder of Refs.~\cite{PhysRevLett.129.200602,agrawal2023observingquantummeasurementcollapse}. However, the utility of the more general perspective is that it allows one to treat general symmetries. In what follows, we discuss the case where the dynamics are strongly symmetric under the nonabelian group $U(1)\rtimes \mathbb{Z}_2$. 

We take the ``noisy decoding'' perspective: we generate measurement records using true quantum evolution, then use a tensor-network based simulation to infer the charge. We compare this tensor-network classifier to the optimal classifier based on full quantum evolution. 
Specifically we consider qubits, whose basis states are $|0\rangle$ and $|1\rangle$, on a chain of length $L$. We numerically generate 20,000 measurement records by running exact dynamics of a hybrid brickwork circuit comprised of strongly symmetric two qubit unitaries occurring with probability $1-p$ and two qubit projective measurements occurring with probability $p$. Half the records are generated by starting from a scrambled $Q=0$ parity symmetric state and the other half from a scrambled $Q=1$ state. The scrambled states are prepared by first starting with a weakly entangled state and then running a brickwork $U(1)\rtimes \mathbb{Z}_2$ symmetric random unitary circuit for $L^2$ time steps. For the $Q=0$ parity symmetric state, we scramble the state, 
$|\psi_{0,+}\rangle \propto \bigotimes_{j=1}^{L/2}|s\rangle_{2j-1,2j}$ where $|s\rangle= \frac{1}{\sqrt{2}}(|10\rangle+|01\rangle)$ 
, and for $Q=1$ we scramble the state $|\psi_{1}\rangle \propto (|11\rangle_{1,2}+|00\rangle_{1,2})\bigotimes_{j=2}^{L/2}|s\rangle_{2j-1,2j}$.  The unitary gates are parameterized as,
\begin{equation}
U_{j,j+1} = P_{j,j+1}^{1}+e^{i \theta_{j,j+1}}P^{a}_{j,j+1}+e^{i\phi_{j,j+1}}P^{s}_{j,j+1},
\end{equation}
where $\theta_{j,j+1},\phi_{j,j+1}$ are uniformly distributed random angles, $P^{1}=|00\rangle\langle00|+|11\rangle\langle11|,P^{s} =|s\rangle\langle s|$, and $P^{a}=|a\rangle\langle a|$ with $|a\rangle = \frac{1}{\sqrt{2}}(|10\rangle-|01\rangle)$. After this scrambling stage, the hybrid brickwork circuit is ran for $L$ time steps where each step consists of gates acting on odd bonds followed by even bonds and the two qubit measurements correspond to acting with $P^{1}$, $P^{s}$, or $P^{a}$ according to the Born rule. We remark that these measurements can be thought of as measuring the energy of a two-site XXZ Hamiltonian, which distinguishes the different irreps of $U(1)\rtimes \mathbb{Z}_2$. After generating these records, one then feeds them into the optimal or noisy decoder to determine the most likely charge.

For the optimal decoder, the same dynamics used to generate a measurement record associated with charge $Q$ is applied to the same initial state, $|\psi_{Q}\rangle$, as well as another scrambled state $|\psi_{Q'}\rangle$ with definite charge $Q'\neq Q$ prepared using the scrambling circuit used to prepare $|\psi_Q\rangle$. After $L$ time steps, one then computes the posterior probability of finding the correct charge associated with $\mathbf{m}$, denoted by $P_{\rm corr}\equiv P(Q_{\rm corr}|\mathbf{m}) = P(\mathbf{m}|Q_{\rm corr})/(P(\mathbf{m}|Q_{\rm wrong})+P(\mathbf{m}|Q_{\rm corr}))$, where $P(\mathbf{m}|Q) = \| |\psi_Q\rangle_{\mathbf{m}} \|^2$, to determine which charge is more likely. 
The procedure is similar for the noisy decoder except one replaces all two qubit unitaries with the following strongly $U(1)\rtimes \mathbb{Z}_2$ symmetric channels,
\begin{equation}
T_{j,j+1}(\cdot) = P_{j,j+1}^1 \cdot P_{j,j+1}^1+P_{j,j+1}^s \cdot P_{j,j+1}^s+P_{j,j+1}^a \cdot P_{j,j+1}^a.
\end{equation}
We remark that one could also use channels with tunable dephasing strength. In that case, we also expect that one should be able to approach the optimal $p_\#$ without encountering a complexity transition.
This replacement means that scrambling is performed with a brickwork circuit comprised of strongly symmetric channels and the resulting scrambled states correspond to maximally mixed states, denoted as $\sigma_Q$ within a given charge subspace. After the scrambling step, one runs hybrid dynamics in a brickwork circuit fashion comprised of the two qubit channels, $T_{j,j+1}$, and projectively measuring according to the measurement record $\mathbf{m}$ from the exact dynamics. One computes the posterior probabilities in the same fashion but now using $P(\mathbf{m}|Q) = \text{tr}(\sigma^{\mathbf{m}}_{Q}(t))$. Observe that instead of computing the posteriors by evolving $\sigma_Q$ one can instead evolve $\mathds{1}$ via the dual channel, i.e. the same channel and measurements ran in reverse, which is only weakly entangled compared to the maximally mixed state $\sigma_Q$ whose bond dimension scales linearly with $L$. As a result, the posterior can be computed as $P(\mathbf{m}|Q) = \text{tr}(\sigma_Q (\mathds{1}(t))^{\mathbf{m}})$.  We numerically run these dynamics using ITensor's~\cite{ITensor,ITensor-r0.3} $U(1)$ conserving version of the time evolving block decimation (TEBD)~\cite{PhysRevLett.91.147902,SCHOLLWOCK201196} algorithm with a fixed truncation error of $\epsilon=10^{-12}$.\par 
In Fig.~\ref{fig:mainfig}, we show the accuracy for the optimal and noisy decoders, defined as the expectation value that the predicted charge (corresponding to the larger of the two posteriors) corresponds to the correct charge. One sees the optimal decoder succeeds above a certain threshold near $p\sim .2$ consistent with sharpening occurring. Importantly, the tensor-network method also clearly has a threshold above which it deterministically succeeds as indicated by the crossing near $p\sim .35$. One can also see this in the distribution of $P_{\rm corr}$---shown in Fig.~\ref{fig:mainfig}(b)---where it skew towards $P_{\rm corr}>1/2$ for larger $p$. Moreover as can be seen in Fig.~\ref{fig:mainfig}(c), in both the fuzzy and sharp phases, its complexity scales polynomially with system size: the required bond dimension scales as $\chi \sim L^{\alpha}$ with $\alpha \approx 2.5$.

\begin{figure*}[!t]
    \centering
    \includegraphics[width=.85\textwidth]{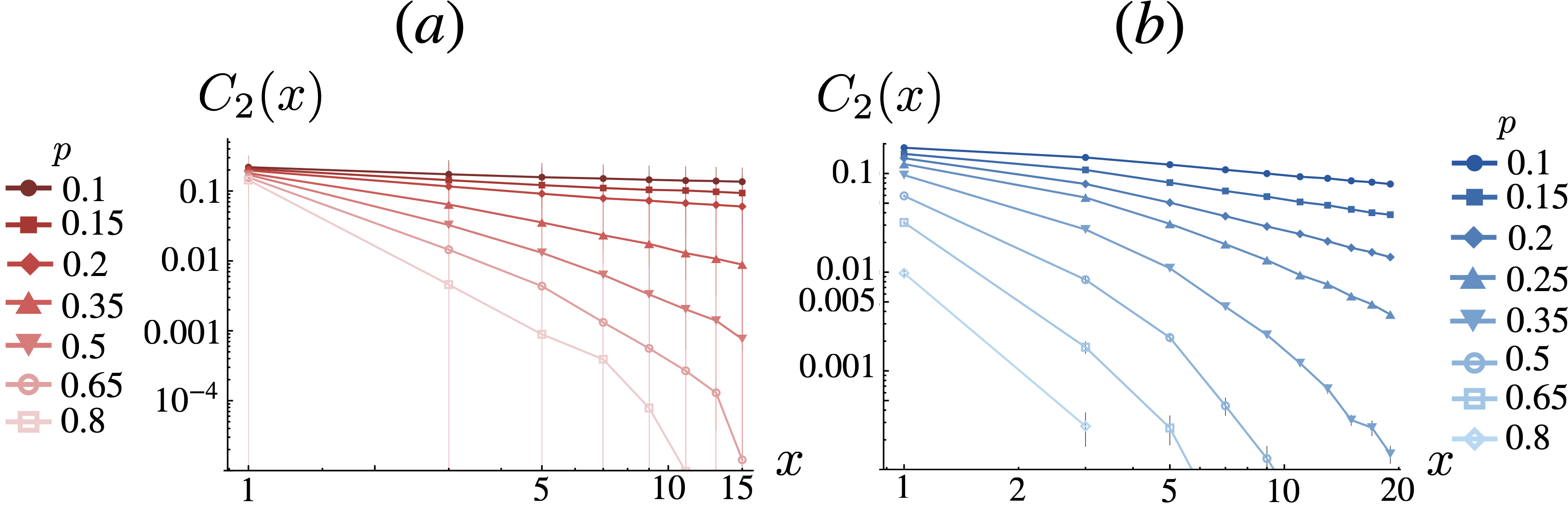}
    \caption{{\bf R\'enyi-2 Correlator .} The R\'enyi-2 correlator for the (a) $U(1)\rtimes \mathbb{Z}_2$ noisy decoder with $L=40$ and (b) $U(1)$ noisy decoder with $L=60$ on a log-log scale evaluated at $t=L$. Both show clear power-law decay for small measurement values and eventually transition to exponential decay for large measurement values. Numerical data is averaged over $10^4$ measurement trajectories simulated with a maximum bond dimension of $\chi=256$. Error bars are the size of markers for the $U(1)$ data. Absent points for $p=0.8$ are a result of correlations decaying too quickly to resolve with the numerical precision used.
     }
    \label{fig:swssbfig}
\end{figure*}

\section{Spontaneous strong-to-weak symmetry breaking (SW-SSB)}

At late enough times, $\sigma^{\mathbf{m}}$ has a definite value of the symmetry charges. Nevertheless, the local structure of this state differs between the sharp and fuzzy phases. This distinction was already pointed out in Ref.~\onlinecite{PhysRevLett.129.120604}: in the fuzzy phase the charge variance in a subsystem of size $\ell$ scales as $\log \ell$, while in the sharp phase it is constant. More generally, the fuzzy phase can be seen as exhibiting spontaneous strong-to-weak symmetry breaking (SW-SSB)~\cite{You24weaksym, Wang24strtowksym}. Heuristically, SW-SSB is said to occur in a state $\rho$ that is globally strongly symmetric if inserting local charge fluctuations in $\rho$ does not lead to a distinguishable state: i.e., if the density matrices $\rho$ and $\hat \rho_x \equiv O^\dagger_x O_0 \rho O^\dagger_0 O_x$ cannot be perfectly distinguished in the limit $x \to \infty$, where $O_i$ is a local operator that is charged under the global strong symmetry. The standard information-theoretic measure of whether two states are distinguishable is the fidelity, $\mathcal{F}(\rho, \sigma) = \left[\mathrm{Tr}(\sqrt{\sqrt{\rho} \sigma \sqrt{\rho}}) \right]^2$. Thus, SW-SSB is defined through the limit $\lim_{x \to \infty} \mathcal{F}(\rho, \hat \rho_x) \neq 0$. In practice, it is often more convenient to use the R\'enyi-2 proxy for this quantity, $\mathcal{C}_2(x) \equiv \mathcal{C}_2(\rho, \hat \rho_x) = \mathrm{Tr}(\rho \hat \rho_x)/\mathrm{Tr}(\rho^2)$. When the states $\rho, \hat \rho_x$ are nearly indistinguishable, $\mathcal{C}_2$ is close to unity; however, like other R\'enyi proxies, in principle it might not accurately detect the onset of SW-SSB. 

The correlator $\mathcal{C}_2(x)$ for typical individual trajectories can be computed in the field theory of charge-sharpening~\cite{PhysRevLett.129.120604} for a $U(1)$ charge. In the limit that is tractable within field theory, all quantum gates are marginalized over~\footnote{An equivalent formulation is that the qubits are coupled to auxiliary charge-neutral, many-level qudits on every site.}, and the density matrix is diagonal in the computational basis at all times, and $\sigma^{\mathbf{m}} = \sum_{\vec{s}} p_{\vec{s}}|\vec{s}\rangle\langle \vec{s}|$ can be mapped to a classical probability distribution $P_{\mathbf{m}}$, which can be written as a ``state'' $| P_{\mathbf{m}} \rrangle = \sum_{\vec s} p_{\vec s} | \vec s \rrangle$, with normalization condition  $\llangle1|P_{\mathbf{m}} \rrangle = 1$ and $| 1 \rrangle = \sum_{\vec s}  | \vec s \rrangle$. In this language, $\mathcal{C}_2(x) \equiv \llangle P_{\mathbf{m}} | \sigma^+(0) \sigma^-(x) | P_{\mathbf{m}} \rrangle / \llangle P_{\mathbf{m}} | P_{\mathbf{m}} \rrangle$, where $\sigma^+(i)$ acts on the probability distribution by adding a particle at site $i$. This correlator can be computed explicitly using the replica field theory language of Ref.~\onlinecite{PhysRevLett.129.120604}. At low measurement rate in one dimension, $\mathcal{C}_2(x)$ exhibits algebraic decay, with a continuously varying exponent, so strict SW-SSB is absent at any finite measurement rate. For stronger measurement rates, in the sharpening phase, SW-SSB is absent and $\mathcal{C}_2$ decays exponentially at large distances. These expectations are borne out by small-size numerics on the noisy decoder (Fig.~\ref{fig:swssbfig} ) In higher dimensions, within the fuzzy phase, $\mathcal{C}_2(x)$ exhibits true long-range order. We expect that similar results can be derived for $\mathcal{F}(\rho, \hat \rho_x)$ by defining this quantity as an appropriate replica limit, but we defer this to future work.

We note that SW-SSB occurs, not just in the conditional state along individual trajectories, but also in the ``system plus measuring apparatus'' density matrix $\Phi \equiv \sum_{\mathbf{m}} p_{\mathbf{m}} \sigma^{\mathbf{m}} \otimes \ket{\mathbf{m}}\bra{\mathbf{m}}$. By orthonormality of the basis $\{ \ket{\mathbf{m}} \}$, we have $\mathcal{C}_2 (\Phi, \hat \Phi_x) = \sum_{\mathbf{m}} p_{\mathbf{m}}^2 \mathrm{Tr}(\sigma^{\mathbf{m}} \hat \sigma^{\mathbf{m}}_x) / \sum_{\mathbf{m}} p_{\mathbf{m}}^2 \mathrm{Tr}((\sigma^{\mathbf{m}})^2)$. 
Thus, if $\lim_{x \to \infty} \mathrm{Tr}(\sigma^{\mathbf{m}} \hat \sigma^{\mathbf{m}}_x) > 0$ for the most probable histories $\mathbf{m}$, it follows that $\Phi$ exhibits SW-SSB. In the field-theory limit of the fuzzy phase, we have already seen that the r.h.s. is nonzero for typical trajectories, so $\Phi$ exhibits SW-SSB in the fuzzy phase. Heuristically, this observation means that the measurement record $\mathbf{m}$ does not pin down the local charge distribution: for any configuration inside a small subregion of size $\ell$, there are $O(\ell)$ configurations with comparable probability in which a charge has been moved far away. On entropic grounds, therefore, SW-SSB is incompatible with a ``sharp'' phase in which the charge variance inside a region remains $O(1)$ as $\ell \to \infty$. 

Away from the limit of perfect dephasing, our calculation of the correlator $\mathcal{C}_2$ is not controlled. It is interesting to consider the purely unitary limit with no dephasing at all. In that limit, a naive extension of our logic would yield the conclusion that the conditional state is $\ket{\psi_{\mathbf{m}}}$, so the fidelity correlator $\mathcal{C}(x) = |\langle \psi_{\mathbf{m}} | \sigma^+_x \sigma^-_0 |\psi_{\mathbf{m}} \rangle|^2$. Since $\ket{\psi_{\mathbf{m}}}$ is a volume-law entangled state, it is implausible for this simple pure-state correlation function to remain nonzero (and in fact, for a model with large auxiliary qudits on every site, one can show that this pure-state correlator is identically zero). Since any noise is expected to lead to states that are essentially classical at long distances, a natural conjecture is that the pure-state limit is singular, and for any nonzero level of dephasing one retains SW-SSB in the mixed state. It would be interesting to perform numerical tests of SW-SSB away from this strong-noise limit.

\section{Discussion}

In this paper we have illustrated how charge-learning/sharpening transitions can be generalized to noisy dynamics and mixed states. We have illustrated how replacing unitary dynamics with noisy dynamics preserves the existence of such transitions, while making them classically simulable. We introduced and employed the strategy of adding noise as a way of making a tradeoff between accuracy and computational efficiency. Finally, we discussed the relation between charge sharpening transitions and spontaneous strong-to-weak symmetry breaking (SW-SSB). Our results provoke many interesting questions for future work, especially about the computational complexity of simulating monitored noisy quantum dynamics. Naively one might expect that measurements do not add complexity to an already easy problem; however, the explicit algorithms for this problem rely on Pauli string expansions that are not valid at any finite measurement rate. An important question for future work is, therefore, to settle the complexity of monitored noisy dynamics, both in cases where the zero-noise limit is known to be easy and in cases (like that of $SU(2)$ symmetry) where it is not.

\begin{acknowledgments}
S.G. thanks Matthew Fisher, David Huse, Sagar Vijay, and Cenke Xu for insightful discussions. R.V. acknowledge partial support from the US Department of Energy, Office of Science, Basic Energy Sciences, under award No. DE-SC0023999. 
A.P. was supported by the Natural Sciences and Engineering Research Council of Canada
(NSERC). This research was supported in part by grant NSF PHY2309135 to the Kavli Institute for Theoretical Physics
(KITP). S.G. was supported by NSF Award No. OMA-2326767.

\end{acknowledgments}

\bibliography{MIPT}

\begin{thebibliography}{95}%
\makeatletter
\providecommand \@ifxundefined [1]{%
 \@ifx{#1\undefined}
}%
\providecommand \@ifnum [1]{%
 \ifnum #1\expandafter \@firstoftwo
 \else \expandafter \@secondoftwo
 \fi
}%
\providecommand \@ifx [1]{%
 \ifx #1\expandafter \@firstoftwo
 \else \expandafter \@secondoftwo
 \fi
}%
\providecommand \natexlab [1]{#1}%
\providecommand \enquote  [1]{``#1''}%
\providecommand \bibnamefont  [1]{#1}%
\providecommand \bibfnamefont [1]{#1}%
\providecommand \citenamefont [1]{#1}%
\providecommand \href@noop [0]{\@secondoftwo}%
\providecommand \href [0]{\begingroup \@sanitize@url \@href}%
\providecommand \@href[1]{\@@startlink{#1}\@@href}%
\providecommand \@@href[1]{\endgroup#1\@@endlink}%
\providecommand \@sanitize@url [0]{\catcode `\\12\catcode `\$12\catcode `\&12\catcode `\#12\catcode `\^12\catcode `\_12\catcode `\%12\relax}%
\providecommand \@@startlink[1]{}%
\providecommand \@@endlink[0]{}%
\providecommand \url  [0]{\begingroup\@sanitize@url \@url }%
\providecommand \@url [1]{\endgroup\@href {#1}{\urlprefix }}%
\providecommand \urlprefix  [0]{URL }%
\providecommand \Eprint [0]{\href }%
\providecommand \doibase [0]{http://dx.doi.org/}%
\providecommand \selectlanguage [0]{\@gobble}%
\providecommand \bibinfo  [0]{\@secondoftwo}%
\providecommand \bibfield  [0]{\@secondoftwo}%
\providecommand \translation [1]{[#1]}%
\providecommand \BibitemOpen [0]{}%
\providecommand \bibitemStop [0]{}%
\providecommand \bibitemNoStop [0]{.\EOS\space}%
\providecommand \EOS [0]{\spacefactor3000\relax}%
\providecommand \BibitemShut  [1]{\csname bibitem#1\endcsname}%
\let\auto@bib@innerbib\@empty
\bibitem [{\citenamefont {Skinner}\ \emph {et~al.}(2019)\citenamefont {Skinner}, \citenamefont {Ruhman},\ and\ \citenamefont {Nahum}}]{PhysRevX.9.031009}%
  \BibitemOpen
  \bibfield  {author} {\bibinfo {author} {\bibfnamefont {B.}~\bibnamefont {Skinner}}, \bibinfo {author} {\bibfnamefont {J.}~\bibnamefont {Ruhman}}, \ and\ \bibinfo {author} {\bibfnamefont {A.}~\bibnamefont {Nahum}},\ }\href {\doibase 10.1103/PhysRevX.9.031009} {\bibfield  {journal} {\bibinfo  {journal} {Phys. Rev. X}\ }\textbf {\bibinfo {volume} {9}},\ \bibinfo {pages} {031009} (\bibinfo {year} {2019})}\BibitemShut {NoStop}%
\bibitem [{\citenamefont {Li}\ \emph {et~al.}(2018)\citenamefont {Li}, \citenamefont {Chen},\ and\ \citenamefont {Fisher}}]{PhysRevB.98.205136}%
  \BibitemOpen
  \bibfield  {author} {\bibinfo {author} {\bibfnamefont {Y.}~\bibnamefont {Li}}, \bibinfo {author} {\bibfnamefont {X.}~\bibnamefont {Chen}}, \ and\ \bibinfo {author} {\bibfnamefont {M.~P.~A.}\ \bibnamefont {Fisher}},\ }\href {\doibase 10.1103/PhysRevB.98.205136} {\bibfield  {journal} {\bibinfo  {journal} {Phys. Rev. B}\ }\textbf {\bibinfo {volume} {98}},\ \bibinfo {pages} {205136} (\bibinfo {year} {2018})}\BibitemShut {NoStop}%
\bibitem [{\citenamefont {Potter}\ and\ \citenamefont {Vasseur}(2022)}]{Potter_2022}%
  \BibitemOpen
  \bibfield  {author} {\bibinfo {author} {\bibfnamefont {A.~C.}\ \bibnamefont {Potter}}\ and\ \bibinfo {author} {\bibfnamefont {R.}~\bibnamefont {Vasseur}},\ }\enquote {\bibinfo {title} {Entanglement dynamics in hybrid quantum circuits},}\ in\ \href {\doibase 10.1007/978-3-031-03998-0_9} {\emph {\bibinfo {booktitle} {Entanglement in Spin Chains}}}\ (\bibinfo  {publisher} {Springer International Publishing},\ \bibinfo {year} {2022})\ p.\ \bibinfo {pages} {211–249}\BibitemShut {NoStop}%
\bibitem [{\citenamefont {Fisher}\ \emph {et~al.}(2023)\citenamefont {Fisher}, \citenamefont {Khemani}, \citenamefont {Nahum},\ and\ \citenamefont {Vijay}}]{annurev:/content/journals/10.1146/annurev-conmatphys-031720-030658}%
  \BibitemOpen
  \bibfield  {author} {\bibinfo {author} {\bibfnamefont {M.~P.}\ \bibnamefont {Fisher}}, \bibinfo {author} {\bibfnamefont {V.}~\bibnamefont {Khemani}}, \bibinfo {author} {\bibfnamefont {A.}~\bibnamefont {Nahum}}, \ and\ \bibinfo {author} {\bibfnamefont {S.}~\bibnamefont {Vijay}},\ }\href {\doibase https://doi.org/10.1146/annurev-conmatphys-031720-030658} {\bibfield  {journal} {\bibinfo  {journal} {Annual Review of Condensed Matter Physics}\ }\textbf {\bibinfo {volume} {14}},\ \bibinfo {pages} {335} (\bibinfo {year} {2023})}\BibitemShut {NoStop}%
\bibitem [{\citenamefont {Gullans}\ and\ \citenamefont {Huse}(2020{\natexlab{a}})}]{PhysRevX.10.041020}%
  \BibitemOpen
  \bibfield  {author} {\bibinfo {author} {\bibfnamefont {M.~J.}\ \bibnamefont {Gullans}}\ and\ \bibinfo {author} {\bibfnamefont {D.~A.}\ \bibnamefont {Huse}},\ }\href {\doibase 10.1103/PhysRevX.10.041020} {\bibfield  {journal} {\bibinfo  {journal} {Phys. Rev. X}\ }\textbf {\bibinfo {volume} {10}},\ \bibinfo {pages} {041020} (\bibinfo {year} {2020}{\natexlab{a}})}\BibitemShut {NoStop}%
\bibitem [{\citenamefont {Choi}\ \emph {et~al.}(2020)\citenamefont {Choi}, \citenamefont {Bao}, \citenamefont {Qi},\ and\ \citenamefont {Altman}}]{PhysRevLett.125.030505}%
  \BibitemOpen
  \bibfield  {author} {\bibinfo {author} {\bibfnamefont {S.}~\bibnamefont {Choi}}, \bibinfo {author} {\bibfnamefont {Y.}~\bibnamefont {Bao}}, \bibinfo {author} {\bibfnamefont {X.-L.}\ \bibnamefont {Qi}}, \ and\ \bibinfo {author} {\bibfnamefont {E.}~\bibnamefont {Altman}},\ }\href {\doibase 10.1103/PhysRevLett.125.030505} {\bibfield  {journal} {\bibinfo  {journal} {Phys. Rev. Lett.}\ }\textbf {\bibinfo {volume} {125}},\ \bibinfo {pages} {030505} (\bibinfo {year} {2020})}\BibitemShut {NoStop}%
\bibitem [{\citenamefont {Li}\ and\ \citenamefont {Fisher}(2021)}]{PhysRevB.103.104306}%
  \BibitemOpen
  \bibfield  {author} {\bibinfo {author} {\bibfnamefont {Y.}~\bibnamefont {Li}}\ and\ \bibinfo {author} {\bibfnamefont {M.~P.~A.}\ \bibnamefont {Fisher}},\ }\href {\doibase 10.1103/PhysRevB.103.104306} {\bibfield  {journal} {\bibinfo  {journal} {Phys. Rev. B}\ }\textbf {\bibinfo {volume} {103}},\ \bibinfo {pages} {104306} (\bibinfo {year} {2021})}\BibitemShut {NoStop}%
\bibitem [{\citenamefont {Fan}\ \emph {et~al.}(2021)\citenamefont {Fan}, \citenamefont {Vijay}, \citenamefont {Vishwanath},\ and\ \citenamefont {You}}]{PhysRevB.103.174309}%
  \BibitemOpen
  \bibfield  {author} {\bibinfo {author} {\bibfnamefont {R.}~\bibnamefont {Fan}}, \bibinfo {author} {\bibfnamefont {S.}~\bibnamefont {Vijay}}, \bibinfo {author} {\bibfnamefont {A.}~\bibnamefont {Vishwanath}}, \ and\ \bibinfo {author} {\bibfnamefont {Y.-Z.}\ \bibnamefont {You}},\ }\href {\doibase 10.1103/PhysRevB.103.174309} {\bibfield  {journal} {\bibinfo  {journal} {Phys. Rev. B}\ }\textbf {\bibinfo {volume} {103}},\ \bibinfo {pages} {174309} (\bibinfo {year} {2021})}\BibitemShut {NoStop}%
\bibitem [{\citenamefont {Li}\ \emph {et~al.}(2019)\citenamefont {Li}, \citenamefont {Chen},\ and\ \citenamefont {Fisher}}]{PhysRevB.100.134306}%
  \BibitemOpen
  \bibfield  {author} {\bibinfo {author} {\bibfnamefont {Y.}~\bibnamefont {Li}}, \bibinfo {author} {\bibfnamefont {X.}~\bibnamefont {Chen}}, \ and\ \bibinfo {author} {\bibfnamefont {M.~P.~A.}\ \bibnamefont {Fisher}},\ }\href {\doibase 10.1103/PhysRevB.100.134306} {\bibfield  {journal} {\bibinfo  {journal} {Phys. Rev. B}\ }\textbf {\bibinfo {volume} {100}},\ \bibinfo {pages} {134306} (\bibinfo {year} {2019})}\BibitemShut {NoStop}%
\bibitem [{\citenamefont {Chan}\ \emph {et~al.}(2019)\citenamefont {Chan}, \citenamefont {Nandkishore}, \citenamefont {Pretko},\ and\ \citenamefont {Smith}}]{PhysRevB.99.224307}%
  \BibitemOpen
  \bibfield  {author} {\bibinfo {author} {\bibfnamefont {A.}~\bibnamefont {Chan}}, \bibinfo {author} {\bibfnamefont {R.~M.}\ \bibnamefont {Nandkishore}}, \bibinfo {author} {\bibfnamefont {M.}~\bibnamefont {Pretko}}, \ and\ \bibinfo {author} {\bibfnamefont {G.}~\bibnamefont {Smith}},\ }\href {\doibase 10.1103/PhysRevB.99.224307} {\bibfield  {journal} {\bibinfo  {journal} {Phys. Rev. B}\ }\textbf {\bibinfo {volume} {99}},\ \bibinfo {pages} {224307} (\bibinfo {year} {2019})}\BibitemShut {NoStop}%
\bibitem [{\citenamefont {Li}\ \emph {et~al.}(2021)\citenamefont {Li}, \citenamefont {Chen}, \citenamefont {Ludwig},\ and\ \citenamefont {Fisher}}]{PhysRevB.104.104305}%
  \BibitemOpen
  \bibfield  {author} {\bibinfo {author} {\bibfnamefont {Y.}~\bibnamefont {Li}}, \bibinfo {author} {\bibfnamefont {X.}~\bibnamefont {Chen}}, \bibinfo {author} {\bibfnamefont {A.~W.~W.}\ \bibnamefont {Ludwig}}, \ and\ \bibinfo {author} {\bibfnamefont {M.~P.~A.}\ \bibnamefont {Fisher}},\ }\href {\doibase 10.1103/PhysRevB.104.104305} {\bibfield  {journal} {\bibinfo  {journal} {Phys. Rev. B}\ }\textbf {\bibinfo {volume} {104}},\ \bibinfo {pages} {104305} (\bibinfo {year} {2021})}\BibitemShut {NoStop}%
\bibitem [{\citenamefont {Szyniszewski}\ \emph {et~al.}(2019)\citenamefont {Szyniszewski}, \citenamefont {Romito},\ and\ \citenamefont {Schomerus}}]{PhysRevB.100.064204}%
  \BibitemOpen
  \bibfield  {author} {\bibinfo {author} {\bibfnamefont {M.}~\bibnamefont {Szyniszewski}}, \bibinfo {author} {\bibfnamefont {A.}~\bibnamefont {Romito}}, \ and\ \bibinfo {author} {\bibfnamefont {H.}~\bibnamefont {Schomerus}},\ }\href {\doibase 10.1103/PhysRevB.100.064204} {\bibfield  {journal} {\bibinfo  {journal} {Phys. Rev. B}\ }\textbf {\bibinfo {volume} {100}},\ \bibinfo {pages} {064204} (\bibinfo {year} {2019})}\BibitemShut {NoStop}%
\bibitem [{\citenamefont {Gullans}\ and\ \citenamefont {Huse}(2020{\natexlab{b}})}]{PhysRevLett.125.070606}%
  \BibitemOpen
  \bibfield  {author} {\bibinfo {author} {\bibfnamefont {M.~J.}\ \bibnamefont {Gullans}}\ and\ \bibinfo {author} {\bibfnamefont {D.~A.}\ \bibnamefont {Huse}},\ }\href {\doibase 10.1103/PhysRevLett.125.070606} {\bibfield  {journal} {\bibinfo  {journal} {Phys. Rev. Lett.}\ }\textbf {\bibinfo {volume} {125}},\ \bibinfo {pages} {070606} (\bibinfo {year} {2020}{\natexlab{b}})}\BibitemShut {NoStop}%
\bibitem [{\citenamefont {Zabalo}\ \emph {et~al.}(2020)\citenamefont {Zabalo}, \citenamefont {Gullans}, \citenamefont {Wilson}, \citenamefont {Gopalakrishnan}, \citenamefont {Huse},\ and\ \citenamefont {Pixley}}]{PhysRevB.101.060301}%
  \BibitemOpen
  \bibfield  {author} {\bibinfo {author} {\bibfnamefont {A.}~\bibnamefont {Zabalo}}, \bibinfo {author} {\bibfnamefont {M.~J.}\ \bibnamefont {Gullans}}, \bibinfo {author} {\bibfnamefont {J.~H.}\ \bibnamefont {Wilson}}, \bibinfo {author} {\bibfnamefont {S.}~\bibnamefont {Gopalakrishnan}}, \bibinfo {author} {\bibfnamefont {D.~A.}\ \bibnamefont {Huse}}, \ and\ \bibinfo {author} {\bibfnamefont {J.~H.}\ \bibnamefont {Pixley}},\ }\href {\doibase 10.1103/PhysRevB.101.060301} {\bibfield  {journal} {\bibinfo  {journal} {Phys. Rev. B}\ }\textbf {\bibinfo {volume} {101}},\ \bibinfo {pages} {060301} (\bibinfo {year} {2020})}\BibitemShut {NoStop}%
\bibitem [{\citenamefont {Ippoliti}\ \emph {et~al.}(2021)\citenamefont {Ippoliti}, \citenamefont {Gullans}, \citenamefont {Gopalakrishnan}, \citenamefont {Huse},\ and\ \citenamefont {Khemani}}]{PhysRevX.11.011030}%
  \BibitemOpen
  \bibfield  {author} {\bibinfo {author} {\bibfnamefont {M.}~\bibnamefont {Ippoliti}}, \bibinfo {author} {\bibfnamefont {M.~J.}\ \bibnamefont {Gullans}}, \bibinfo {author} {\bibfnamefont {S.}~\bibnamefont {Gopalakrishnan}}, \bibinfo {author} {\bibfnamefont {D.~A.}\ \bibnamefont {Huse}}, \ and\ \bibinfo {author} {\bibfnamefont {V.}~\bibnamefont {Khemani}},\ }\href {\doibase 10.1103/PhysRevX.11.011030} {\bibfield  {journal} {\bibinfo  {journal} {Phys. Rev. X}\ }\textbf {\bibinfo {volume} {11}},\ \bibinfo {pages} {011030} (\bibinfo {year} {2021})}\BibitemShut {NoStop}%
\bibitem [{\citenamefont {Lavasani}\ \emph {et~al.}(2021)\citenamefont {Lavasani}, \citenamefont {Alavirad},\ and\ \citenamefont {Barkeshli}}]{lavasani_measurement-induced_2021}%
  \BibitemOpen
  \bibfield  {author} {\bibinfo {author} {\bibfnamefont {A.}~\bibnamefont {Lavasani}}, \bibinfo {author} {\bibfnamefont {Y.}~\bibnamefont {Alavirad}}, \ and\ \bibinfo {author} {\bibfnamefont {M.}~\bibnamefont {Barkeshli}},\ }\href {\doibase 10.1038/s41567-020-01112-z} {\bibfield  {journal} {\bibinfo  {journal} {Nature Physics}\ }\textbf {\bibinfo {volume} {17}},\ \bibinfo {pages} {342} (\bibinfo {year} {2021})}\BibitemShut {NoStop}%
\bibitem [{\citenamefont {Sang}\ and\ \citenamefont {Hsieh}(2021)}]{PhysRevResearch.3.023200}%
  \BibitemOpen
  \bibfield  {author} {\bibinfo {author} {\bibfnamefont {S.}~\bibnamefont {Sang}}\ and\ \bibinfo {author} {\bibfnamefont {T.~H.}\ \bibnamefont {Hsieh}},\ }\href {\doibase 10.1103/PhysRevResearch.3.023200} {\bibfield  {journal} {\bibinfo  {journal} {Phys. Rev. Res.}\ }\textbf {\bibinfo {volume} {3}},\ \bibinfo {pages} {023200} (\bibinfo {year} {2021})}\BibitemShut {NoStop}%
\bibitem [{\citenamefont {Tang}\ and\ \citenamefont {Zhu}(2020)}]{PhysRevResearch.2.013022}%
  \BibitemOpen
  \bibfield  {author} {\bibinfo {author} {\bibfnamefont {Q.}~\bibnamefont {Tang}}\ and\ \bibinfo {author} {\bibfnamefont {W.}~\bibnamefont {Zhu}},\ }\href {\doibase 10.1103/PhysRevResearch.2.013022} {\bibfield  {journal} {\bibinfo  {journal} {Phys. Rev. Res.}\ }\textbf {\bibinfo {volume} {2}},\ \bibinfo {pages} {013022} (\bibinfo {year} {2020})}\BibitemShut {NoStop}%
\bibitem [{\citenamefont {Turkeshi}\ \emph {et~al.}(2020)\citenamefont {Turkeshi}, \citenamefont {Fazio},\ and\ \citenamefont {Dalmonte}}]{PhysRevB.102.014315}%
  \BibitemOpen
  \bibfield  {author} {\bibinfo {author} {\bibfnamefont {X.}~\bibnamefont {Turkeshi}}, \bibinfo {author} {\bibfnamefont {R.}~\bibnamefont {Fazio}}, \ and\ \bibinfo {author} {\bibfnamefont {M.}~\bibnamefont {Dalmonte}},\ }\href {\doibase 10.1103/PhysRevB.102.014315} {\bibfield  {journal} {\bibinfo  {journal} {Phys. Rev. B}\ }\textbf {\bibinfo {volume} {102}},\ \bibinfo {pages} {014315} (\bibinfo {year} {2020})}\BibitemShut {NoStop}%
\bibitem [{\citenamefont {Fuji}\ and\ \citenamefont {Ashida}(2020)}]{PhysRevB.102.054302}%
  \BibitemOpen
  \bibfield  {author} {\bibinfo {author} {\bibfnamefont {Y.}~\bibnamefont {Fuji}}\ and\ \bibinfo {author} {\bibfnamefont {Y.}~\bibnamefont {Ashida}},\ }\href {\doibase 10.1103/PhysRevB.102.054302} {\bibfield  {journal} {\bibinfo  {journal} {Phys. Rev. B}\ }\textbf {\bibinfo {volume} {102}},\ \bibinfo {pages} {054302} (\bibinfo {year} {2020})}\BibitemShut {NoStop}%
\bibitem [{\citenamefont {Lunt}\ \emph {et~al.}(2021)\citenamefont {Lunt}, \citenamefont {Szyniszewski},\ and\ \citenamefont {Pal}}]{PhysRevB.104.155111}%
  \BibitemOpen
  \bibfield  {author} {\bibinfo {author} {\bibfnamefont {O.}~\bibnamefont {Lunt}}, \bibinfo {author} {\bibfnamefont {M.}~\bibnamefont {Szyniszewski}}, \ and\ \bibinfo {author} {\bibfnamefont {A.}~\bibnamefont {Pal}},\ }\href {\doibase 10.1103/PhysRevB.104.155111} {\bibfield  {journal} {\bibinfo  {journal} {Phys. Rev. B}\ }\textbf {\bibinfo {volume} {104}},\ \bibinfo {pages} {155111} (\bibinfo {year} {2021})}\BibitemShut {NoStop}%
\bibitem [{\citenamefont {Turkeshi}\ \emph {et~al.}(2021)\citenamefont {Turkeshi}, \citenamefont {Biella}, \citenamefont {Fazio}, \citenamefont {Dalmonte},\ and\ \citenamefont {Schir\'o}}]{PhysRevB.103.224210}%
  \BibitemOpen
  \bibfield  {author} {\bibinfo {author} {\bibfnamefont {X.}~\bibnamefont {Turkeshi}}, \bibinfo {author} {\bibfnamefont {A.}~\bibnamefont {Biella}}, \bibinfo {author} {\bibfnamefont {R.}~\bibnamefont {Fazio}}, \bibinfo {author} {\bibfnamefont {M.}~\bibnamefont {Dalmonte}}, \ and\ \bibinfo {author} {\bibfnamefont {M.}~\bibnamefont {Schir\'o}},\ }\href {\doibase 10.1103/PhysRevB.103.224210} {\bibfield  {journal} {\bibinfo  {journal} {Phys. Rev. B}\ }\textbf {\bibinfo {volume} {103}},\ \bibinfo {pages} {224210} (\bibinfo {year} {2021})}\BibitemShut {NoStop}%
\bibitem [{\citenamefont {Ippoliti}\ and\ \citenamefont {Khemani}(2021)}]{PhysRevLett.126.060501}%
  \BibitemOpen
  \bibfield  {author} {\bibinfo {author} {\bibfnamefont {M.}~\bibnamefont {Ippoliti}}\ and\ \bibinfo {author} {\bibfnamefont {V.}~\bibnamefont {Khemani}},\ }\href {\doibase 10.1103/PhysRevLett.126.060501} {\bibfield  {journal} {\bibinfo  {journal} {Phys. Rev. Lett.}\ }\textbf {\bibinfo {volume} {126}},\ \bibinfo {pages} {060501} (\bibinfo {year} {2021})}\BibitemShut {NoStop}%
\bibitem [{\citenamefont {Lu}\ and\ \citenamefont {Grover}(2021)}]{PRXQuantum.2.040319}%
  \BibitemOpen
  \bibfield  {author} {\bibinfo {author} {\bibfnamefont {T.-C.}\ \bibnamefont {Lu}}\ and\ \bibinfo {author} {\bibfnamefont {T.}~\bibnamefont {Grover}},\ }\href {\doibase 10.1103/PRXQuantum.2.040319} {\bibfield  {journal} {\bibinfo  {journal} {PRX Quantum}\ }\textbf {\bibinfo {volume} {2}},\ \bibinfo {pages} {040319} (\bibinfo {year} {2021})}\BibitemShut {NoStop}%
\bibitem [{\citenamefont {Jian}\ \emph {et~al.}(2022)\citenamefont {Jian}, \citenamefont {Bauer}, \citenamefont {Keselman},\ and\ \citenamefont {Ludwig}}]{PhysRevB.106.134206}%
  \BibitemOpen
  \bibfield  {author} {\bibinfo {author} {\bibfnamefont {C.-M.}\ \bibnamefont {Jian}}, \bibinfo {author} {\bibfnamefont {B.}~\bibnamefont {Bauer}}, \bibinfo {author} {\bibfnamefont {A.}~\bibnamefont {Keselman}}, \ and\ \bibinfo {author} {\bibfnamefont {A.~W.~W.}\ \bibnamefont {Ludwig}},\ }\href {\doibase 10.1103/PhysRevB.106.134206} {\bibfield  {journal} {\bibinfo  {journal} {Phys. Rev. B}\ }\textbf {\bibinfo {volume} {106}},\ \bibinfo {pages} {134206} (\bibinfo {year} {2022})}\BibitemShut {NoStop}%
\bibitem [{\citenamefont {Gopalakrishnan}\ and\ \citenamefont {Gullans}(2021)}]{PhysRevLett.126.170503}%
  \BibitemOpen
  \bibfield  {author} {\bibinfo {author} {\bibfnamefont {S.}~\bibnamefont {Gopalakrishnan}}\ and\ \bibinfo {author} {\bibfnamefont {M.~J.}\ \bibnamefont {Gullans}},\ }\href {\doibase 10.1103/PhysRevLett.126.170503} {\bibfield  {journal} {\bibinfo  {journal} {Phys. Rev. Lett.}\ }\textbf {\bibinfo {volume} {126}},\ \bibinfo {pages} {170503} (\bibinfo {year} {2021})}\BibitemShut {NoStop}%
\bibitem [{\citenamefont {Turkeshi}(2022)}]{PhysRevB.106.144313}%
  \BibitemOpen
  \bibfield  {author} {\bibinfo {author} {\bibfnamefont {X.}~\bibnamefont {Turkeshi}},\ }\href {\doibase 10.1103/PhysRevB.106.144313} {\bibfield  {journal} {\bibinfo  {journal} {Phys. Rev. B}\ }\textbf {\bibinfo {volume} {106}},\ \bibinfo {pages} {144313} (\bibinfo {year} {2022})}\BibitemShut {NoStop}%
\bibitem [{\citenamefont {Block}\ \emph {et~al.}(2022)\citenamefont {Block}, \citenamefont {Bao}, \citenamefont {Choi}, \citenamefont {Altman},\ and\ \citenamefont {Yao}}]{PhysRevLett.128.010604}%
  \BibitemOpen
  \bibfield  {author} {\bibinfo {author} {\bibfnamefont {M.}~\bibnamefont {Block}}, \bibinfo {author} {\bibfnamefont {Y.}~\bibnamefont {Bao}}, \bibinfo {author} {\bibfnamefont {S.}~\bibnamefont {Choi}}, \bibinfo {author} {\bibfnamefont {E.}~\bibnamefont {Altman}}, \ and\ \bibinfo {author} {\bibfnamefont {N.~Y.}\ \bibnamefont {Yao}},\ }\href {\doibase 10.1103/PhysRevLett.128.010604} {\bibfield  {journal} {\bibinfo  {journal} {Phys. Rev. Lett.}\ }\textbf {\bibinfo {volume} {128}},\ \bibinfo {pages} {010604} (\bibinfo {year} {2022})}\BibitemShut {NoStop}%
\bibitem [{\citenamefont {Bentsen}\ \emph {et~al.}(2021)\citenamefont {Bentsen}, \citenamefont {Sahu},\ and\ \citenamefont {Swingle}}]{PhysRevB.104.094304}%
  \BibitemOpen
  \bibfield  {author} {\bibinfo {author} {\bibfnamefont {G.~S.}\ \bibnamefont {Bentsen}}, \bibinfo {author} {\bibfnamefont {S.}~\bibnamefont {Sahu}}, \ and\ \bibinfo {author} {\bibfnamefont {B.}~\bibnamefont {Swingle}},\ }\href {\doibase 10.1103/PhysRevB.104.094304} {\bibfield  {journal} {\bibinfo  {journal} {Phys. Rev. B}\ }\textbf {\bibinfo {volume} {104}},\ \bibinfo {pages} {094304} (\bibinfo {year} {2021})}\BibitemShut {NoStop}%
\bibitem [{\citenamefont {Zabalo}\ \emph {et~al.}(2022)\citenamefont {Zabalo}, \citenamefont {Gullans}, \citenamefont {Wilson}, \citenamefont {Vasseur}, \citenamefont {Ludwig}, \citenamefont {Gopalakrishnan}, \citenamefont {Huse},\ and\ \citenamefont {Pixley}}]{PhysRevLett.128.050602}%
  \BibitemOpen
  \bibfield  {author} {\bibinfo {author} {\bibfnamefont {A.}~\bibnamefont {Zabalo}}, \bibinfo {author} {\bibfnamefont {M.~J.}\ \bibnamefont {Gullans}}, \bibinfo {author} {\bibfnamefont {J.~H.}\ \bibnamefont {Wilson}}, \bibinfo {author} {\bibfnamefont {R.}~\bibnamefont {Vasseur}}, \bibinfo {author} {\bibfnamefont {A.~W.~W.}\ \bibnamefont {Ludwig}}, \bibinfo {author} {\bibfnamefont {S.}~\bibnamefont {Gopalakrishnan}}, \bibinfo {author} {\bibfnamefont {D.~A.}\ \bibnamefont {Huse}}, \ and\ \bibinfo {author} {\bibfnamefont {J.~H.}\ \bibnamefont {Pixley}},\ }\href {\doibase 10.1103/PhysRevLett.128.050602} {\bibfield  {journal} {\bibinfo  {journal} {Phys. Rev. Lett.}\ }\textbf {\bibinfo {volume} {128}},\ \bibinfo {pages} {050602} (\bibinfo {year} {2022})}\BibitemShut {NoStop}%
\bibitem [{\citenamefont {Li}\ and\ \citenamefont {Fisher}(2023)}]{PhysRevB.108.214302}%
  \BibitemOpen
  \bibfield  {author} {\bibinfo {author} {\bibfnamefont {Y.}~\bibnamefont {Li}}\ and\ \bibinfo {author} {\bibfnamefont {M.~P.~A.}\ \bibnamefont {Fisher}},\ }\href {\doibase 10.1103/PhysRevB.108.214302} {\bibfield  {journal} {\bibinfo  {journal} {Phys. Rev. B}\ }\textbf {\bibinfo {volume} {108}},\ \bibinfo {pages} {214302} (\bibinfo {year} {2023})}\BibitemShut {NoStop}%
\bibitem [{\citenamefont {Jian}\ \emph {et~al.}(2021)\citenamefont {Jian}, \citenamefont {Liu}, \citenamefont {Chen}, \citenamefont {Swingle},\ and\ \citenamefont {Zhang}}]{jian2021quantumerroremergentmagnetic}%
  \BibitemOpen
  \bibfield  {author} {\bibinfo {author} {\bibfnamefont {S.-K.}\ \bibnamefont {Jian}}, \bibinfo {author} {\bibfnamefont {C.}~\bibnamefont {Liu}}, \bibinfo {author} {\bibfnamefont {X.}~\bibnamefont {Chen}}, \bibinfo {author} {\bibfnamefont {B.}~\bibnamefont {Swingle}}, \ and\ \bibinfo {author} {\bibfnamefont {P.}~\bibnamefont {Zhang}},\ }\href {https://arxiv.org/abs/2106.09635} {\enquote {\bibinfo {title} {Quantum error as an emergent magnetic field},}\ } (\bibinfo {year} {2021}),\ \Eprint {http://arxiv.org/abs/2106.09635} {arXiv:2106.09635 [quant-ph]} \BibitemShut {NoStop}%
\bibitem [{\citenamefont {Doggen}\ \emph {et~al.}(2022)\citenamefont {Doggen}, \citenamefont {Gefen}, \citenamefont {Gornyi}, \citenamefont {Mirlin},\ and\ \citenamefont {Polyakov}}]{PhysRevResearch.4.023146}%
  \BibitemOpen
  \bibfield  {author} {\bibinfo {author} {\bibfnamefont {E.~V.~H.}\ \bibnamefont {Doggen}}, \bibinfo {author} {\bibfnamefont {Y.}~\bibnamefont {Gefen}}, \bibinfo {author} {\bibfnamefont {I.~V.}\ \bibnamefont {Gornyi}}, \bibinfo {author} {\bibfnamefont {A.~D.}\ \bibnamefont {Mirlin}}, \ and\ \bibinfo {author} {\bibfnamefont {D.~G.}\ \bibnamefont {Polyakov}},\ }\href {\doibase 10.1103/PhysRevResearch.4.023146} {\bibfield  {journal} {\bibinfo  {journal} {Phys. Rev. Res.}\ }\textbf {\bibinfo {volume} {4}},\ \bibinfo {pages} {023146} (\bibinfo {year} {2022})}\BibitemShut {NoStop}%
\bibitem [{\citenamefont {Sierant}\ and\ \citenamefont {Turkeshi}(2022)}]{PhysRevLett.128.130605}%
  \BibitemOpen
  \bibfield  {author} {\bibinfo {author} {\bibfnamefont {P.}~\bibnamefont {Sierant}}\ and\ \bibinfo {author} {\bibfnamefont {X.}~\bibnamefont {Turkeshi}},\ }\href {\doibase 10.1103/PhysRevLett.128.130605} {\bibfield  {journal} {\bibinfo  {journal} {Phys. Rev. Lett.}\ }\textbf {\bibinfo {volume} {128}},\ \bibinfo {pages} {130605} (\bibinfo {year} {2022})}\BibitemShut {NoStop}%
\bibitem [{\citenamefont {Sierant}\ \emph {et~al.}(2022)\citenamefont {Sierant}, \citenamefont {Chiriac{\`{o}}}, \citenamefont {Surace}, \citenamefont {Sharma}, \citenamefont {Turkeshi}, \citenamefont {Dalmonte}, \citenamefont {Fazio},\ and\ \citenamefont {Pagano}}]{Sierant2022dissipativefloquet}%
  \BibitemOpen
  \bibfield  {author} {\bibinfo {author} {\bibfnamefont {P.}~\bibnamefont {Sierant}}, \bibinfo {author} {\bibfnamefont {G.}~\bibnamefont {Chiriac{\`{o}}}}, \bibinfo {author} {\bibfnamefont {F.~M.}\ \bibnamefont {Surace}}, \bibinfo {author} {\bibfnamefont {S.}~\bibnamefont {Sharma}}, \bibinfo {author} {\bibfnamefont {X.}~\bibnamefont {Turkeshi}}, \bibinfo {author} {\bibfnamefont {M.}~\bibnamefont {Dalmonte}}, \bibinfo {author} {\bibfnamefont {R.}~\bibnamefont {Fazio}}, \ and\ \bibinfo {author} {\bibfnamefont {G.}~\bibnamefont {Pagano}},\ }\href {\doibase 10.22331/q-2022-02-02-638} {\bibfield  {journal} {\bibinfo  {journal} {{Quantum}}\ }\textbf {\bibinfo {volume} {6}},\ \bibinfo {pages} {638} (\bibinfo {year} {2022})}\BibitemShut {NoStop}%
\bibitem [{\citenamefont {Sharma}\ \emph {et~al.}(2022)\citenamefont {Sharma}, \citenamefont {Turkeshi}, \citenamefont {Fazio},\ and\ \citenamefont {Dalmonte}}]{10.21468/SciPostPhysCore.5.2.023}%
  \BibitemOpen
  \bibfield  {author} {\bibinfo {author} {\bibfnamefont {S.}~\bibnamefont {Sharma}}, \bibinfo {author} {\bibfnamefont {X.}~\bibnamefont {Turkeshi}}, \bibinfo {author} {\bibfnamefont {R.}~\bibnamefont {Fazio}}, \ and\ \bibinfo {author} {\bibfnamefont {M.}~\bibnamefont {Dalmonte}},\ }\href {\doibase 10.21468/SciPostPhysCore.5.2.023} {\bibfield  {journal} {\bibinfo  {journal} {SciPost Phys. Core}\ }\textbf {\bibinfo {volume} {5}},\ \bibinfo {pages} {023} (\bibinfo {year} {2022})}\BibitemShut {NoStop}%
\bibitem [{\citenamefont {Li}\ \emph {et~al.}(2023{\natexlab{a}})\citenamefont {Li}, \citenamefont {Zou}, \citenamefont {Glorioso}, \citenamefont {Altman},\ and\ \citenamefont {Fisher}}]{PhysRevLett.130.220404}%
  \BibitemOpen
  \bibfield  {author} {\bibinfo {author} {\bibfnamefont {Y.}~\bibnamefont {Li}}, \bibinfo {author} {\bibfnamefont {Y.}~\bibnamefont {Zou}}, \bibinfo {author} {\bibfnamefont {P.}~\bibnamefont {Glorioso}}, \bibinfo {author} {\bibfnamefont {E.}~\bibnamefont {Altman}}, \ and\ \bibinfo {author} {\bibfnamefont {M.~P.~A.}\ \bibnamefont {Fisher}},\ }\href {\doibase 10.1103/PhysRevLett.130.220404} {\bibfield  {journal} {\bibinfo  {journal} {Phys. Rev. Lett.}\ }\textbf {\bibinfo {volume} {130}},\ \bibinfo {pages} {220404} (\bibinfo {year} {2023}{\natexlab{a}})}\BibitemShut {NoStop}%
\bibitem [{\citenamefont {Negari}\ \emph {et~al.}(2024)\citenamefont {Negari}, \citenamefont {Sahu},\ and\ \citenamefont {Hsieh}}]{PhysRevB.109.125148}%
  \BibitemOpen
  \bibfield  {author} {\bibinfo {author} {\bibfnamefont {A.-R.}\ \bibnamefont {Negari}}, \bibinfo {author} {\bibfnamefont {S.}~\bibnamefont {Sahu}}, \ and\ \bibinfo {author} {\bibfnamefont {T.~H.}\ \bibnamefont {Hsieh}},\ }\href {\doibase 10.1103/PhysRevB.109.125148} {\bibfield  {journal} {\bibinfo  {journal} {Phys. Rev. B}\ }\textbf {\bibinfo {volume} {109}},\ \bibinfo {pages} {125148} (\bibinfo {year} {2024})}\BibitemShut {NoStop}%
\bibitem [{\citenamefont {Cecile}\ \emph {et~al.}(2024)\citenamefont {Cecile}, \citenamefont {L\'oio},\ and\ \citenamefont {De~Nardis}}]{PhysRevResearch.6.033220}%
  \BibitemOpen
  \bibfield  {author} {\bibinfo {author} {\bibfnamefont {G.}~\bibnamefont {Cecile}}, \bibinfo {author} {\bibfnamefont {H.}~\bibnamefont {L\'oio}}, \ and\ \bibinfo {author} {\bibfnamefont {J.}~\bibnamefont {De~Nardis}},\ }\href {\doibase 10.1103/PhysRevResearch.6.033220} {\bibfield  {journal} {\bibinfo  {journal} {Phys. Rev. Res.}\ }\textbf {\bibinfo {volume} {6}},\ \bibinfo {pages} {033220} (\bibinfo {year} {2024})}\BibitemShut {NoStop}%
\bibitem [{\citenamefont {Feng}\ \emph {et~al.}(2023{\natexlab{a}})\citenamefont {Feng}, \citenamefont {Skinner},\ and\ \citenamefont {Nahum}}]{PRXQuantum.4.030333}%
  \BibitemOpen
  \bibfield  {author} {\bibinfo {author} {\bibfnamefont {X.}~\bibnamefont {Feng}}, \bibinfo {author} {\bibfnamefont {B.}~\bibnamefont {Skinner}}, \ and\ \bibinfo {author} {\bibfnamefont {A.}~\bibnamefont {Nahum}},\ }\href {\doibase 10.1103/PRXQuantum.4.030333} {\bibfield  {journal} {\bibinfo  {journal} {PRX Quantum}\ }\textbf {\bibinfo {volume} {4}},\ \bibinfo {pages} {030333} (\bibinfo {year} {2023}{\natexlab{a}})}\BibitemShut {NoStop}%
\bibitem [{\citenamefont {Heinrich}\ and\ \citenamefont {Chen}(2024)}]{PhysRevB.110.064309}%
  \BibitemOpen
  \bibfield  {author} {\bibinfo {author} {\bibfnamefont {E.}~\bibnamefont {Heinrich}}\ and\ \bibinfo {author} {\bibfnamefont {X.}~\bibnamefont {Chen}},\ }\href {\doibase 10.1103/PhysRevB.110.064309} {\bibfield  {journal} {\bibinfo  {journal} {Phys. Rev. B}\ }\textbf {\bibinfo {volume} {110}},\ \bibinfo {pages} {064309} (\bibinfo {year} {2024})}\BibitemShut {NoStop}%
\bibitem [{\citenamefont {Aziz}\ \emph {et~al.}(2024)\citenamefont {Aziz}, \citenamefont {Chakraborty},\ and\ \citenamefont {Pixley}}]{PhysRevB.110.064301}%
  \BibitemOpen
  \bibfield  {author} {\bibinfo {author} {\bibfnamefont {K.}~\bibnamefont {Aziz}}, \bibinfo {author} {\bibfnamefont {A.}~\bibnamefont {Chakraborty}}, \ and\ \bibinfo {author} {\bibfnamefont {J.~H.}\ \bibnamefont {Pixley}},\ }\href {\doibase 10.1103/PhysRevB.110.064301} {\bibfield  {journal} {\bibinfo  {journal} {Phys. Rev. B}\ }\textbf {\bibinfo {volume} {110}},\ \bibinfo {pages} {064301} (\bibinfo {year} {2024})}\BibitemShut {NoStop}%
\bibitem [{\citenamefont {Jin}\ and\ \citenamefont {Martin}(2024)}]{PhysRevB.110.L060202}%
  \BibitemOpen
  \bibfield  {author} {\bibinfo {author} {\bibfnamefont {T.}~\bibnamefont {Jin}}\ and\ \bibinfo {author} {\bibfnamefont {D.~G.}\ \bibnamefont {Martin}},\ }\href {\doibase 10.1103/PhysRevB.110.L060202} {\bibfield  {journal} {\bibinfo  {journal} {Phys. Rev. B}\ }\textbf {\bibinfo {volume} {110}},\ \bibinfo {pages} {L060202} (\bibinfo {year} {2024})}\BibitemShut {NoStop}%
\bibitem [{\citenamefont {Paviglianiti}\ and\ \citenamefont {Silva}(2023)}]{PhysRevB.108.184302}%
  \BibitemOpen
  \bibfield  {author} {\bibinfo {author} {\bibfnamefont {A.}~\bibnamefont {Paviglianiti}}\ and\ \bibinfo {author} {\bibfnamefont {A.}~\bibnamefont {Silva}},\ }\href {\doibase 10.1103/PhysRevB.108.184302} {\bibfield  {journal} {\bibinfo  {journal} {Phys. Rev. B}\ }\textbf {\bibinfo {volume} {108}},\ \bibinfo {pages} {184302} (\bibinfo {year} {2023})}\BibitemShut {NoStop}%
\bibitem [{\citenamefont {Zabalo}\ \emph {et~al.}(2023)\citenamefont {Zabalo}, \citenamefont {Wilson}, \citenamefont {Gullans}, \citenamefont {Vasseur}, \citenamefont {Gopalakrishnan}, \citenamefont {Huse},\ and\ \citenamefont {Pixley}}]{PhysRevB.107.L220204}%
  \BibitemOpen
  \bibfield  {author} {\bibinfo {author} {\bibfnamefont {A.}~\bibnamefont {Zabalo}}, \bibinfo {author} {\bibfnamefont {J.~H.}\ \bibnamefont {Wilson}}, \bibinfo {author} {\bibfnamefont {M.~J.}\ \bibnamefont {Gullans}}, \bibinfo {author} {\bibfnamefont {R.}~\bibnamefont {Vasseur}}, \bibinfo {author} {\bibfnamefont {S.}~\bibnamefont {Gopalakrishnan}}, \bibinfo {author} {\bibfnamefont {D.~A.}\ \bibnamefont {Huse}}, \ and\ \bibinfo {author} {\bibfnamefont {J.~H.}\ \bibnamefont {Pixley}},\ }\href {\doibase 10.1103/PhysRevB.107.L220204} {\bibfield  {journal} {\bibinfo  {journal} {Phys. Rev. B}\ }\textbf {\bibinfo {volume} {107}},\ \bibinfo {pages} {L220204} (\bibinfo {year} {2023})}\BibitemShut {NoStop}%
\bibitem [{\citenamefont {Shkolnik}\ \emph {et~al.}(2023)\citenamefont {Shkolnik}, \citenamefont {Zabalo}, \citenamefont {Vasseur}, \citenamefont {Huse}, \citenamefont {Pixley},\ and\ \citenamefont {Gazit}}]{PhysRevB.108.184204}%
  \BibitemOpen
  \bibfield  {author} {\bibinfo {author} {\bibfnamefont {G.}~\bibnamefont {Shkolnik}}, \bibinfo {author} {\bibfnamefont {A.}~\bibnamefont {Zabalo}}, \bibinfo {author} {\bibfnamefont {R.}~\bibnamefont {Vasseur}}, \bibinfo {author} {\bibfnamefont {D.~A.}\ \bibnamefont {Huse}}, \bibinfo {author} {\bibfnamefont {J.~H.}\ \bibnamefont {Pixley}}, \ and\ \bibinfo {author} {\bibfnamefont {S.}~\bibnamefont {Gazit}},\ }\href {\doibase 10.1103/PhysRevB.108.184204} {\bibfield  {journal} {\bibinfo  {journal} {Phys. Rev. B}\ }\textbf {\bibinfo {volume} {108}},\ \bibinfo {pages} {184204} (\bibinfo {year} {2023})}\BibitemShut {NoStop}%
\bibitem [{\citenamefont {Kumar}\ \emph {et~al.}(2024)\citenamefont {Kumar}, \citenamefont {Aziz}, \citenamefont {Chakraborty}, \citenamefont {Ludwig}, \citenamefont {Gopalakrishnan}, \citenamefont {Pixley},\ and\ \citenamefont {Vasseur}}]{PhysRevB.109.014303}%
  \BibitemOpen
  \bibfield  {author} {\bibinfo {author} {\bibfnamefont {A.}~\bibnamefont {Kumar}}, \bibinfo {author} {\bibfnamefont {K.}~\bibnamefont {Aziz}}, \bibinfo {author} {\bibfnamefont {A.}~\bibnamefont {Chakraborty}}, \bibinfo {author} {\bibfnamefont {A.~W.~W.}\ \bibnamefont {Ludwig}}, \bibinfo {author} {\bibfnamefont {S.}~\bibnamefont {Gopalakrishnan}}, \bibinfo {author} {\bibfnamefont {J.~H.}\ \bibnamefont {Pixley}}, \ and\ \bibinfo {author} {\bibfnamefont {R.}~\bibnamefont {Vasseur}},\ }\href {\doibase 10.1103/PhysRevB.109.014303} {\bibfield  {journal} {\bibinfo  {journal} {Phys. Rev. B}\ }\textbf {\bibinfo {volume} {109}},\ \bibinfo {pages} {014303} (\bibinfo {year} {2024})}\BibitemShut {NoStop}%
\bibitem [{\citenamefont {Liu}\ \emph {et~al.}(2023)\citenamefont {Liu}, \citenamefont {Li}, \citenamefont {Zhang}, \citenamefont {Jian},\ and\ \citenamefont {Yao}}]{PhysRevB.107.L201113}%
  \BibitemOpen
  \bibfield  {author} {\bibinfo {author} {\bibfnamefont {S.}~\bibnamefont {Liu}}, \bibinfo {author} {\bibfnamefont {M.-R.}\ \bibnamefont {Li}}, \bibinfo {author} {\bibfnamefont {S.-X.}\ \bibnamefont {Zhang}}, \bibinfo {author} {\bibfnamefont {S.-K.}\ \bibnamefont {Jian}}, \ and\ \bibinfo {author} {\bibfnamefont {H.}~\bibnamefont {Yao}},\ }\href {\doibase 10.1103/PhysRevB.107.L201113} {\bibfield  {journal} {\bibinfo  {journal} {Phys. Rev. B}\ }\textbf {\bibinfo {volume} {107}},\ \bibinfo {pages} {L201113} (\bibinfo {year} {2023})}\BibitemShut {NoStop}%
\bibitem [{\citenamefont {Le~Gal}\ \emph {et~al.}(2024)\citenamefont {Le~Gal}, \citenamefont {Turkeshi},\ and\ \citenamefont {Schir\`o}}]{PRXQuantum.5.030329}%
  \BibitemOpen
  \bibfield  {author} {\bibinfo {author} {\bibfnamefont {Y.}~\bibnamefont {Le~Gal}}, \bibinfo {author} {\bibfnamefont {X.}~\bibnamefont {Turkeshi}}, \ and\ \bibinfo {author} {\bibfnamefont {M.}~\bibnamefont {Schir\`o}},\ }\href {\doibase 10.1103/PRXQuantum.5.030329} {\bibfield  {journal} {\bibinfo  {journal} {PRX Quantum}\ }\textbf {\bibinfo {volume} {5}},\ \bibinfo {pages} {030329} (\bibinfo {year} {2024})}\BibitemShut {NoStop}%
\bibitem [{\citenamefont {Pan}\ \emph {et~al.}(2024)\citenamefont {Pan}, \citenamefont {Ganeshan}, \citenamefont {Iadecola}, \citenamefont {Wilson},\ and\ \citenamefont {Pixley}}]{PhysRevB.110.054308}%
  \BibitemOpen
  \bibfield  {author} {\bibinfo {author} {\bibfnamefont {H.}~\bibnamefont {Pan}}, \bibinfo {author} {\bibfnamefont {S.}~\bibnamefont {Ganeshan}}, \bibinfo {author} {\bibfnamefont {T.}~\bibnamefont {Iadecola}}, \bibinfo {author} {\bibfnamefont {J.~H.}\ \bibnamefont {Wilson}}, \ and\ \bibinfo {author} {\bibfnamefont {J.~H.}\ \bibnamefont {Pixley}},\ }\href {\doibase 10.1103/PhysRevB.110.054308} {\bibfield  {journal} {\bibinfo  {journal} {Phys. Rev. B}\ }\textbf {\bibinfo {volume} {110}},\ \bibinfo {pages} {054308} (\bibinfo {year} {2024})}\BibitemShut {NoStop}%
\bibitem [{\citenamefont {Lumia}\ \emph {et~al.}(2024)\citenamefont {Lumia}, \citenamefont {Tirrito}, \citenamefont {Fazio},\ and\ \citenamefont {Collura}}]{PhysRevResearch.6.023176}%
  \BibitemOpen
  \bibfield  {author} {\bibinfo {author} {\bibfnamefont {L.}~\bibnamefont {Lumia}}, \bibinfo {author} {\bibfnamefont {E.}~\bibnamefont {Tirrito}}, \bibinfo {author} {\bibfnamefont {R.}~\bibnamefont {Fazio}}, \ and\ \bibinfo {author} {\bibfnamefont {M.}~\bibnamefont {Collura}},\ }\href {\doibase 10.1103/PhysRevResearch.6.023176} {\bibfield  {journal} {\bibinfo  {journal} {Phys. Rev. Res.}\ }\textbf {\bibinfo {volume} {6}},\ \bibinfo {pages} {023176} (\bibinfo {year} {2024})}\BibitemShut {NoStop}%
\bibitem [{\citenamefont {Sierant}\ and\ \citenamefont {Turkeshi}(2023)}]{PhysRevLett.130.120402}%
  \BibitemOpen
  \bibfield  {author} {\bibinfo {author} {\bibfnamefont {P.}~\bibnamefont {Sierant}}\ and\ \bibinfo {author} {\bibfnamefont {X.}~\bibnamefont {Turkeshi}},\ }\href {\doibase 10.1103/PhysRevLett.130.120402} {\bibfield  {journal} {\bibinfo  {journal} {Phys. Rev. Lett.}\ }\textbf {\bibinfo {volume} {130}},\ \bibinfo {pages} {120402} (\bibinfo {year} {2023})}\BibitemShut {NoStop}%
\bibitem [{\citenamefont {Liu}\ \emph {et~al.}(2024{\natexlab{a}})\citenamefont {Liu}, \citenamefont {Li}, \citenamefont {Zhang},\ and\ \citenamefont {Jian}}]{PhysRevLett.132.240402}%
  \BibitemOpen
  \bibfield  {author} {\bibinfo {author} {\bibfnamefont {S.}~\bibnamefont {Liu}}, \bibinfo {author} {\bibfnamefont {M.-R.}\ \bibnamefont {Li}}, \bibinfo {author} {\bibfnamefont {S.-X.}\ \bibnamefont {Zhang}}, \ and\ \bibinfo {author} {\bibfnamefont {S.-K.}\ \bibnamefont {Jian}},\ }\href {\doibase 10.1103/PhysRevLett.132.240402} {\bibfield  {journal} {\bibinfo  {journal} {Phys. Rev. Lett.}\ }\textbf {\bibinfo {volume} {132}},\ \bibinfo {pages} {240402} (\bibinfo {year} {2024}{\natexlab{a}})}\BibitemShut {NoStop}%
\bibitem [{\citenamefont {Granet}\ \emph {et~al.}(2023)\citenamefont {Granet}, \citenamefont {Zhang},\ and\ \citenamefont {Dreyer}}]{PhysRevLett.130.230401}%
  \BibitemOpen
  \bibfield  {author} {\bibinfo {author} {\bibfnamefont {E.}~\bibnamefont {Granet}}, \bibinfo {author} {\bibfnamefont {C.}~\bibnamefont {Zhang}}, \ and\ \bibinfo {author} {\bibfnamefont {H.}~\bibnamefont {Dreyer}},\ }\href {\doibase 10.1103/PhysRevLett.130.230401} {\bibfield  {journal} {\bibinfo  {journal} {Phys. Rev. Lett.}\ }\textbf {\bibinfo {volume} {130}},\ \bibinfo {pages} {230401} (\bibinfo {year} {2023})}\BibitemShut {NoStop}%
\bibitem [{\citenamefont {Yamamoto}\ and\ \citenamefont {Hamazaki}(2023)}]{PhysRevB.107.L220201}%
  \BibitemOpen
  \bibfield  {author} {\bibinfo {author} {\bibfnamefont {K.}~\bibnamefont {Yamamoto}}\ and\ \bibinfo {author} {\bibfnamefont {R.}~\bibnamefont {Hamazaki}},\ }\href {\doibase 10.1103/PhysRevB.107.L220201} {\bibfield  {journal} {\bibinfo  {journal} {Phys. Rev. B}\ }\textbf {\bibinfo {volume} {107}},\ \bibinfo {pages} {L220201} (\bibinfo {year} {2023})}\BibitemShut {NoStop}%
\bibitem [{\citenamefont {Chakraborty}\ \emph {et~al.}(2024)\citenamefont {Chakraborty}, \citenamefont {Chen}, \citenamefont {Zabalo}, \citenamefont {Wilson},\ and\ \citenamefont {Pixley}}]{PhysRevB.110.045135}%
  \BibitemOpen
  \bibfield  {author} {\bibinfo {author} {\bibfnamefont {A.}~\bibnamefont {Chakraborty}}, \bibinfo {author} {\bibfnamefont {K.}~\bibnamefont {Chen}}, \bibinfo {author} {\bibfnamefont {A.}~\bibnamefont {Zabalo}}, \bibinfo {author} {\bibfnamefont {J.~H.}\ \bibnamefont {Wilson}}, \ and\ \bibinfo {author} {\bibfnamefont {J.~H.}\ \bibnamefont {Pixley}},\ }\href {\doibase 10.1103/PhysRevB.110.045135} {\bibfield  {journal} {\bibinfo  {journal} {Phys. Rev. B}\ }\textbf {\bibinfo {volume} {110}},\ \bibinfo {pages} {045135} (\bibinfo {year} {2024})}\BibitemShut {NoStop}%
\bibitem [{\citenamefont {Ravindranath}\ \emph {et~al.}(2023)\citenamefont {Ravindranath}, \citenamefont {Han}, \citenamefont {Yang},\ and\ \citenamefont {Chen}}]{PhysRevB.108.L041103}%
  \BibitemOpen
  \bibfield  {author} {\bibinfo {author} {\bibfnamefont {V.}~\bibnamefont {Ravindranath}}, \bibinfo {author} {\bibfnamefont {Y.}~\bibnamefont {Han}}, \bibinfo {author} {\bibfnamefont {Z.-C.}\ \bibnamefont {Yang}}, \ and\ \bibinfo {author} {\bibfnamefont {X.}~\bibnamefont {Chen}},\ }\href {\doibase 10.1103/PhysRevB.108.L041103} {\bibfield  {journal} {\bibinfo  {journal} {Phys. Rev. B}\ }\textbf {\bibinfo {volume} {108}},\ \bibinfo {pages} {L041103} (\bibinfo {year} {2023})}\BibitemShut {NoStop}%
\bibitem [{\citenamefont {Feng}\ \emph {et~al.}(2023{\natexlab{b}})\citenamefont {Feng}, \citenamefont {Liu}, \citenamefont {Chen},\ and\ \citenamefont {Guo}}]{PhysRevB.107.094309}%
  \BibitemOpen
  \bibfield  {author} {\bibinfo {author} {\bibfnamefont {X.}~\bibnamefont {Feng}}, \bibinfo {author} {\bibfnamefont {S.}~\bibnamefont {Liu}}, \bibinfo {author} {\bibfnamefont {S.}~\bibnamefont {Chen}}, \ and\ \bibinfo {author} {\bibfnamefont {W.}~\bibnamefont {Guo}},\ }\href {\doibase 10.1103/PhysRevB.107.094309} {\bibfield  {journal} {\bibinfo  {journal} {Phys. Rev. B}\ }\textbf {\bibinfo {volume} {107}},\ \bibinfo {pages} {094309} (\bibinfo {year} {2023}{\natexlab{b}})}\BibitemShut {NoStop}%
\bibitem [{\citenamefont {Liu}\ \emph {et~al.}(2024{\natexlab{b}})\citenamefont {Liu}, \citenamefont {Li}, \citenamefont {Zhang}, \citenamefont {Jian},\ and\ \citenamefont {Yao}}]{PhysRevB.110.064323}%
  \BibitemOpen
  \bibfield  {author} {\bibinfo {author} {\bibfnamefont {S.}~\bibnamefont {Liu}}, \bibinfo {author} {\bibfnamefont {M.-R.}\ \bibnamefont {Li}}, \bibinfo {author} {\bibfnamefont {S.-X.}\ \bibnamefont {Zhang}}, \bibinfo {author} {\bibfnamefont {S.-K.}\ \bibnamefont {Jian}}, \ and\ \bibinfo {author} {\bibfnamefont {H.}~\bibnamefont {Yao}},\ }\href {\doibase 10.1103/PhysRevB.110.064323} {\bibfield  {journal} {\bibinfo  {journal} {Phys. Rev. B}\ }\textbf {\bibinfo {volume} {110}},\ \bibinfo {pages} {064323} (\bibinfo {year} {2024}{\natexlab{b}})}\BibitemShut {NoStop}%
\bibitem [{\citenamefont {Oshima}\ and\ \citenamefont {Fuji}(2023)}]{PhysRevB.107.014308}%
  \BibitemOpen
  \bibfield  {author} {\bibinfo {author} {\bibfnamefont {H.}~\bibnamefont {Oshima}}\ and\ \bibinfo {author} {\bibfnamefont {Y.}~\bibnamefont {Fuji}},\ }\href {\doibase 10.1103/PhysRevB.107.014308} {\bibfield  {journal} {\bibinfo  {journal} {Phys. Rev. B}\ }\textbf {\bibinfo {volume} {107}},\ \bibinfo {pages} {014308} (\bibinfo {year} {2023})}\BibitemShut {NoStop}%
\bibitem [{\citenamefont {Piroli}\ \emph {et~al.}(2023)\citenamefont {Piroli}, \citenamefont {Li}, \citenamefont {Vasseur},\ and\ \citenamefont {Nahum}}]{PhysRevB.107.224303}%
  \BibitemOpen
  \bibfield  {author} {\bibinfo {author} {\bibfnamefont {L.}~\bibnamefont {Piroli}}, \bibinfo {author} {\bibfnamefont {Y.}~\bibnamefont {Li}}, \bibinfo {author} {\bibfnamefont {R.}~\bibnamefont {Vasseur}}, \ and\ \bibinfo {author} {\bibfnamefont {A.}~\bibnamefont {Nahum}},\ }\href {\doibase 10.1103/PhysRevB.107.224303} {\bibfield  {journal} {\bibinfo  {journal} {Phys. Rev. B}\ }\textbf {\bibinfo {volume} {107}},\ \bibinfo {pages} {224303} (\bibinfo {year} {2023})}\BibitemShut {NoStop}%
\bibitem [{\citenamefont {Iadecola}\ \emph {et~al.}(2023)\citenamefont {Iadecola}, \citenamefont {Ganeshan}, \citenamefont {Pixley},\ and\ \citenamefont {Wilson}}]{PhysRevLett.131.060403}%
  \BibitemOpen
  \bibfield  {author} {\bibinfo {author} {\bibfnamefont {T.}~\bibnamefont {Iadecola}}, \bibinfo {author} {\bibfnamefont {S.}~\bibnamefont {Ganeshan}}, \bibinfo {author} {\bibfnamefont {J.~H.}\ \bibnamefont {Pixley}}, \ and\ \bibinfo {author} {\bibfnamefont {J.~H.}\ \bibnamefont {Wilson}},\ }\href {\doibase 10.1103/PhysRevLett.131.060403} {\bibfield  {journal} {\bibinfo  {journal} {Phys. Rev. Lett.}\ }\textbf {\bibinfo {volume} {131}},\ \bibinfo {pages} {060403} (\bibinfo {year} {2023})}\BibitemShut {NoStop}%
\bibitem [{\citenamefont {Weinstein}\ \emph {et~al.}(2023)\citenamefont {Weinstein}, \citenamefont {Kelly}, \citenamefont {Marino},\ and\ \citenamefont {Altman}}]{PhysRevLett.131.220404}%
  \BibitemOpen
  \bibfield  {author} {\bibinfo {author} {\bibfnamefont {Z.}~\bibnamefont {Weinstein}}, \bibinfo {author} {\bibfnamefont {S.~P.}\ \bibnamefont {Kelly}}, \bibinfo {author} {\bibfnamefont {J.}~\bibnamefont {Marino}}, \ and\ \bibinfo {author} {\bibfnamefont {E.}~\bibnamefont {Altman}},\ }\href {\doibase 10.1103/PhysRevLett.131.220404} {\bibfield  {journal} {\bibinfo  {journal} {Phys. Rev. Lett.}\ }\textbf {\bibinfo {volume} {131}},\ \bibinfo {pages} {220404} (\bibinfo {year} {2023})}\BibitemShut {NoStop}%
\bibitem [{\citenamefont {Li}\ and\ \citenamefont {Claassen}(2023)}]{PhysRevB.108.104310}%
  \BibitemOpen
  \bibfield  {author} {\bibinfo {author} {\bibfnamefont {Y.}~\bibnamefont {Li}}\ and\ \bibinfo {author} {\bibfnamefont {M.}~\bibnamefont {Claassen}},\ }\href {\doibase 10.1103/PhysRevB.108.104310} {\bibfield  {journal} {\bibinfo  {journal} {Phys. Rev. B}\ }\textbf {\bibinfo {volume} {108}},\ \bibinfo {pages} {104310} (\bibinfo {year} {2023})}\BibitemShut {NoStop}%
\bibitem [{\citenamefont {Szyniszewski}\ \emph {et~al.}(2023)\citenamefont {Szyniszewski}, \citenamefont {Lunt},\ and\ \citenamefont {Pal}}]{PhysRevB.108.165126}%
  \BibitemOpen
  \bibfield  {author} {\bibinfo {author} {\bibfnamefont {M.}~\bibnamefont {Szyniszewski}}, \bibinfo {author} {\bibfnamefont {O.}~\bibnamefont {Lunt}}, \ and\ \bibinfo {author} {\bibfnamefont {A.}~\bibnamefont {Pal}},\ }\href {\doibase 10.1103/PhysRevB.108.165126} {\bibfield  {journal} {\bibinfo  {journal} {Phys. Rev. B}\ }\textbf {\bibinfo {volume} {108}},\ \bibinfo {pages} {165126} (\bibinfo {year} {2023})}\BibitemShut {NoStop}%
\bibitem [{\citenamefont {Nahum}\ and\ \citenamefont {Wiese}(2023)}]{PhysRevB.108.104203}%
  \BibitemOpen
  \bibfield  {author} {\bibinfo {author} {\bibfnamefont {A.}~\bibnamefont {Nahum}}\ and\ \bibinfo {author} {\bibfnamefont {K.~J.}\ \bibnamefont {Wiese}},\ }\href {\doibase 10.1103/PhysRevB.108.104203} {\bibfield  {journal} {\bibinfo  {journal} {Phys. Rev. B}\ }\textbf {\bibinfo {volume} {108}},\ \bibinfo {pages} {104203} (\bibinfo {year} {2023})}\BibitemShut {NoStop}%
\bibitem [{\citenamefont {Merritt}\ and\ \citenamefont {Fidkowski}(2023)}]{PhysRevB.107.064303}%
  \BibitemOpen
  \bibfield  {author} {\bibinfo {author} {\bibfnamefont {J.}~\bibnamefont {Merritt}}\ and\ \bibinfo {author} {\bibfnamefont {L.}~\bibnamefont {Fidkowski}},\ }\href {\doibase 10.1103/PhysRevB.107.064303} {\bibfield  {journal} {\bibinfo  {journal} {Phys. Rev. B}\ }\textbf {\bibinfo {volume} {107}},\ \bibinfo {pages} {064303} (\bibinfo {year} {2023})}\BibitemShut {NoStop}%
\bibitem [{\citenamefont {Li}\ \emph {et~al.}(2023{\natexlab{b}})\citenamefont {Li}, \citenamefont {Vijay},\ and\ \citenamefont {Fisher}}]{Li_2023}%
  \BibitemOpen
  \bibfield  {author} {\bibinfo {author} {\bibfnamefont {Y.}~\bibnamefont {Li}}, \bibinfo {author} {\bibfnamefont {S.}~\bibnamefont {Vijay}}, \ and\ \bibinfo {author} {\bibfnamefont {M.~P.}\ \bibnamefont {Fisher}},\ }\href {\doibase 10.1103/prxquantum.4.010331} {\bibfield  {journal} {\bibinfo  {journal} {PRX Quantum}\ }\textbf {\bibinfo {volume} {4}} (\bibinfo {year} {2023}{\natexlab{b}}),\ 10.1103/prxquantum.4.010331}\BibitemShut {NoStop}%
\bibitem [{\citenamefont {L\'oio}\ \emph {et~al.}(2023)\citenamefont {L\'oio}, \citenamefont {De~Luca}, \citenamefont {De~Nardis},\ and\ \citenamefont {Turkeshi}}]{PhysRevB.108.L020306}%
  \BibitemOpen
  \bibfield  {author} {\bibinfo {author} {\bibfnamefont {H.}~\bibnamefont {L\'oio}}, \bibinfo {author} {\bibfnamefont {A.}~\bibnamefont {De~Luca}}, \bibinfo {author} {\bibfnamefont {J.}~\bibnamefont {De~Nardis}}, \ and\ \bibinfo {author} {\bibfnamefont {X.}~\bibnamefont {Turkeshi}},\ }\href {\doibase 10.1103/PhysRevB.108.L020306} {\bibfield  {journal} {\bibinfo  {journal} {Phys. Rev. B}\ }\textbf {\bibinfo {volume} {108}},\ \bibinfo {pages} {L020306} (\bibinfo {year} {2023})}\BibitemShut {NoStop}%
\bibitem [{\citenamefont {Jian}\ \emph {et~al.}(2023)\citenamefont {Jian}, \citenamefont {Shapourian}, \citenamefont {Bauer},\ and\ \citenamefont {Ludwig}}]{JianShapourianBauerLudwig2023}%
  \BibitemOpen
  \bibfield  {author} {\bibinfo {author} {\bibfnamefont {C.-M.}\ \bibnamefont {Jian}}, \bibinfo {author} {\bibfnamefont {H.}~\bibnamefont {Shapourian}}, \bibinfo {author} {\bibfnamefont {B.}~\bibnamefont {Bauer}}, \ and\ \bibinfo {author} {\bibfnamefont {A.~W.~W.}\ \bibnamefont {Ludwig}},\ }\href {https://arxiv.org/abs/2302.09094} {\enquote {\bibinfo {title} {Measurement-induced entanglement transitions in quantum circuits of non-interacting fermions: Born-rule versus forced measurements},}\ } (\bibinfo {year} {2023}),\ \Eprint {http://arxiv.org/abs/2302.09094} {arXiv:2302.09094 [cond-mat.stat-mech]} \BibitemShut {NoStop}%
\bibitem [{\citenamefont {Fava}\ \emph {et~al.}(2023)\citenamefont {Fava}, \citenamefont {Piroli}, \citenamefont {Swann}, \citenamefont {Bernard},\ and\ \citenamefont {Nahum}}]{PhysRevX.13.041045}%
  \BibitemOpen
  \bibfield  {author} {\bibinfo {author} {\bibfnamefont {M.}~\bibnamefont {Fava}}, \bibinfo {author} {\bibfnamefont {L.}~\bibnamefont {Piroli}}, \bibinfo {author} {\bibfnamefont {T.}~\bibnamefont {Swann}}, \bibinfo {author} {\bibfnamefont {D.}~\bibnamefont {Bernard}}, \ and\ \bibinfo {author} {\bibfnamefont {A.}~\bibnamefont {Nahum}},\ }\href {\doibase 10.1103/PhysRevX.13.041045} {\bibfield  {journal} {\bibinfo  {journal} {Phys. Rev. X}\ }\textbf {\bibinfo {volume} {13}},\ \bibinfo {pages} {041045} (\bibinfo {year} {2023})}\BibitemShut {NoStop}%
\bibitem [{\citenamefont {Sang}\ \emph {et~al.}(2023)\citenamefont {Sang}, \citenamefont {Li}, \citenamefont {Hsieh},\ and\ \citenamefont {Yoshida}}]{PRXQuantum.4.040332}%
  \BibitemOpen
  \bibfield  {author} {\bibinfo {author} {\bibfnamefont {S.}~\bibnamefont {Sang}}, \bibinfo {author} {\bibfnamefont {Z.}~\bibnamefont {Li}}, \bibinfo {author} {\bibfnamefont {T.~H.}\ \bibnamefont {Hsieh}}, \ and\ \bibinfo {author} {\bibfnamefont {B.}~\bibnamefont {Yoshida}},\ }\href {\doibase 10.1103/PRXQuantum.4.040332} {\bibfield  {journal} {\bibinfo  {journal} {PRX Quantum}\ }\textbf {\bibinfo {volume} {4}},\ \bibinfo {pages} {040332} (\bibinfo {year} {2023})}\BibitemShut {NoStop}%
\bibitem [{\citenamefont {Bao}\ \emph {et~al.}(2021)\citenamefont {Bao}, \citenamefont {Choi},\ and\ \citenamefont {Altman}}]{BAO2021168618}%
  \BibitemOpen
  \bibfield  {author} {\bibinfo {author} {\bibfnamefont {Y.}~\bibnamefont {Bao}}, \bibinfo {author} {\bibfnamefont {S.}~\bibnamefont {Choi}}, \ and\ \bibinfo {author} {\bibfnamefont {E.}~\bibnamefont {Altman}},\ }\href {\doibase https://doi.org/10.1016/j.aop.2021.168618} {\bibfield  {journal} {\bibinfo  {journal} {Annals of Physics}\ }\textbf {\bibinfo {volume} {435}},\ \bibinfo {pages} {168618} (\bibinfo {year} {2021})},\ \bibinfo {note} {special issue on Philip W. Anderson}\BibitemShut {NoStop}%
\bibitem [{\citenamefont {Agrawal}\ \emph {et~al.}(2022)\citenamefont {Agrawal}, \citenamefont {Zabalo}, \citenamefont {Chen}, \citenamefont {Wilson}, \citenamefont {Potter}, \citenamefont {Pixley}, \citenamefont {Gopalakrishnan},\ and\ \citenamefont {Vasseur}}]{PhysRevX.12.041002}%
  \BibitemOpen
  \bibfield  {author} {\bibinfo {author} {\bibfnamefont {U.}~\bibnamefont {Agrawal}}, \bibinfo {author} {\bibfnamefont {A.}~\bibnamefont {Zabalo}}, \bibinfo {author} {\bibfnamefont {K.}~\bibnamefont {Chen}}, \bibinfo {author} {\bibfnamefont {J.~H.}\ \bibnamefont {Wilson}}, \bibinfo {author} {\bibfnamefont {A.~C.}\ \bibnamefont {Potter}}, \bibinfo {author} {\bibfnamefont {J.~H.}\ \bibnamefont {Pixley}}, \bibinfo {author} {\bibfnamefont {S.}~\bibnamefont {Gopalakrishnan}}, \ and\ \bibinfo {author} {\bibfnamefont {R.}~\bibnamefont {Vasseur}},\ }\href {\doibase 10.1103/PhysRevX.12.041002} {\bibfield  {journal} {\bibinfo  {journal} {Phys. Rev. X}\ }\textbf {\bibinfo {volume} {12}},\ \bibinfo {pages} {041002} (\bibinfo {year} {2022})}\BibitemShut {NoStop}%
\bibitem [{\citenamefont {Majidy}\ \emph {et~al.}(2023)\citenamefont {Majidy}, \citenamefont {Agrawal}, \citenamefont {Gopalakrishnan}, \citenamefont {Potter}, \citenamefont {Vasseur},\ and\ \citenamefont {Halpern}}]{PhysRevB.108.054307}%
  \BibitemOpen
  \bibfield  {author} {\bibinfo {author} {\bibfnamefont {S.}~\bibnamefont {Majidy}}, \bibinfo {author} {\bibfnamefont {U.}~\bibnamefont {Agrawal}}, \bibinfo {author} {\bibfnamefont {S.}~\bibnamefont {Gopalakrishnan}}, \bibinfo {author} {\bibfnamefont {A.~C.}\ \bibnamefont {Potter}}, \bibinfo {author} {\bibfnamefont {R.}~\bibnamefont {Vasseur}}, \ and\ \bibinfo {author} {\bibfnamefont {N.~Y.}\ \bibnamefont {Halpern}},\ }\href {\doibase 10.1103/PhysRevB.108.054307} {\bibfield  {journal} {\bibinfo  {journal} {Phys. Rev. B}\ }\textbf {\bibinfo {volume} {108}},\ \bibinfo {pages} {054307} (\bibinfo {year} {2023})}\BibitemShut {NoStop}%
\bibitem [{\citenamefont {Guo}\ \emph {et~al.}(2024)\citenamefont {Guo}, \citenamefont {Foster}, \citenamefont {Jian},\ and\ \citenamefont {Ludwig}}]{GuoJianFosterLudwigKeldysh2024}%
  \BibitemOpen
  \bibfield  {author} {\bibinfo {author} {\bibfnamefont {H.}~\bibnamefont {Guo}}, \bibinfo {author} {\bibfnamefont {M.~S.}\ \bibnamefont {Foster}}, \bibinfo {author} {\bibfnamefont {C.-M.}\ \bibnamefont {Jian}}, \ and\ \bibinfo {author} {\bibfnamefont {A.~W.~W.}\ \bibnamefont {Ludwig}},\ }\href {https://arxiv.org/abs/2410.07317} {\enquote {\bibinfo {title} {Field theory of monitored, interacting fermion dynamics with charge conservation},}\ } (\bibinfo {year} {2024}),\ \Eprint {http://arxiv.org/abs/2410.07317} {arXiv:2410.07317 [cond-mat.stat-mech]} \BibitemShut {NoStop}%
\bibitem [{\citenamefont {Poboiko}\ \emph {et~al.}(2024)\citenamefont {Poboiko}, \citenamefont {Pöpperl}, \citenamefont {Gornyi},\ and\ \citenamefont {Mirlin}}]{MirlinGornyiEtAlKeldysh2024}%
  \BibitemOpen
  \bibfield  {author} {\bibinfo {author} {\bibfnamefont {I.}~\bibnamefont {Poboiko}}, \bibinfo {author} {\bibfnamefont {P.}~\bibnamefont {Pöpperl}}, \bibinfo {author} {\bibfnamefont {I.~V.}\ \bibnamefont {Gornyi}}, \ and\ \bibinfo {author} {\bibfnamefont {A.~D.}\ \bibnamefont {Mirlin}},\ }\href {https://arxiv.org/abs/2410.07334} {\enquote {\bibinfo {title} {Measurement-induced transitions for interacting fermions},}\ } (\bibinfo {year} {2024}),\ \Eprint {http://arxiv.org/abs/2410.07334} {arXiv:2410.07334 [quant-ph]} \BibitemShut {NoStop}%
\bibitem [{\citenamefont {Cao}\ \emph {et~al.}(2019)\citenamefont {Cao}, \citenamefont {Tilloy},\ and\ \citenamefont {Luca}}]{10.21468/SciPostPhys.7.2.024}%
  \BibitemOpen
  \bibfield  {author} {\bibinfo {author} {\bibfnamefont {X.}~\bibnamefont {Cao}}, \bibinfo {author} {\bibfnamefont {A.}~\bibnamefont {Tilloy}}, \ and\ \bibinfo {author} {\bibfnamefont {A.~D.}\ \bibnamefont {Luca}},\ }\href {\doibase 10.21468/SciPostPhys.7.2.024} {\bibfield  {journal} {\bibinfo  {journal} {SciPost Phys.}\ }\textbf {\bibinfo {volume} {7}},\ \bibinfo {pages} {024} (\bibinfo {year} {2019})}\BibitemShut {NoStop}%
\bibitem [{\citenamefont {Nahum}\ and\ \citenamefont {Skinner}(2020)}]{PhysRevResearch.2.023288}%
  \BibitemOpen
  \bibfield  {author} {\bibinfo {author} {\bibfnamefont {A.}~\bibnamefont {Nahum}}\ and\ \bibinfo {author} {\bibfnamefont {B.}~\bibnamefont {Skinner}},\ }\href {\doibase 10.1103/PhysRevResearch.2.023288} {\bibfield  {journal} {\bibinfo  {journal} {Phys. Rev. Res.}\ }\textbf {\bibinfo {volume} {2}},\ \bibinfo {pages} {023288} (\bibinfo {year} {2020})}\BibitemShut {NoStop}%
\bibitem [{\citenamefont {Alberton}\ \emph {et~al.}(2021)\citenamefont {Alberton}, \citenamefont {Buchhold},\ and\ \citenamefont {Diehl}}]{PhysRevLett.126.170602}%
  \BibitemOpen
  \bibfield  {author} {\bibinfo {author} {\bibfnamefont {O.}~\bibnamefont {Alberton}}, \bibinfo {author} {\bibfnamefont {M.}~\bibnamefont {Buchhold}}, \ and\ \bibinfo {author} {\bibfnamefont {S.}~\bibnamefont {Diehl}},\ }\href {\doibase 10.1103/PhysRevLett.126.170602} {\bibfield  {journal} {\bibinfo  {journal} {Phys. Rev. Lett.}\ }\textbf {\bibinfo {volume} {126}},\ \bibinfo {pages} {170602} (\bibinfo {year} {2021})}\BibitemShut {NoStop}%
\bibitem [{\citenamefont {Buchhold}\ \emph {et~al.}(2021)\citenamefont {Buchhold}, \citenamefont {Minoguchi}, \citenamefont {Altland},\ and\ \citenamefont {Diehl}}]{PhysRevX.11.041004}%
  \BibitemOpen
  \bibfield  {author} {\bibinfo {author} {\bibfnamefont {M.}~\bibnamefont {Buchhold}}, \bibinfo {author} {\bibfnamefont {Y.}~\bibnamefont {Minoguchi}}, \bibinfo {author} {\bibfnamefont {A.}~\bibnamefont {Altland}}, \ and\ \bibinfo {author} {\bibfnamefont {S.}~\bibnamefont {Diehl}},\ }\href {\doibase 10.1103/PhysRevX.11.041004} {\bibfield  {journal} {\bibinfo  {journal} {Phys. Rev. X}\ }\textbf {\bibinfo {volume} {11}},\ \bibinfo {pages} {041004} (\bibinfo {year} {2021})}\BibitemShut {NoStop}%
\bibitem [{\citenamefont {Poboiko}\ \emph {et~al.}(2023)\citenamefont {Poboiko}, \citenamefont {P\"opperl}, \citenamefont {Gornyi},\ and\ \citenamefont {Mirlin}}]{PhysRevX.13.041046}%
  \BibitemOpen
  \bibfield  {author} {\bibinfo {author} {\bibfnamefont {I.}~\bibnamefont {Poboiko}}, \bibinfo {author} {\bibfnamefont {P.}~\bibnamefont {P\"opperl}}, \bibinfo {author} {\bibfnamefont {I.~V.}\ \bibnamefont {Gornyi}}, \ and\ \bibinfo {author} {\bibfnamefont {A.~D.}\ \bibnamefont {Mirlin}},\ }\href {\doibase 10.1103/PhysRevX.13.041046} {\bibfield  {journal} {\bibinfo  {journal} {Phys. Rev. X}\ }\textbf {\bibinfo {volume} {13}},\ \bibinfo {pages} {041046} (\bibinfo {year} {2023})}\BibitemShut {NoStop}%
\bibitem [{\citenamefont {Barratt}\ \emph {et~al.}(2022{\natexlab{a}})\citenamefont {Barratt}, \citenamefont {Agrawal}, \citenamefont {Potter}, \citenamefont {Gopalakrishnan},\ and\ \citenamefont {Vasseur}}]{PhysRevLett.129.200602}%
  \BibitemOpen
  \bibfield  {author} {\bibinfo {author} {\bibfnamefont {F.}~\bibnamefont {Barratt}}, \bibinfo {author} {\bibfnamefont {U.}~\bibnamefont {Agrawal}}, \bibinfo {author} {\bibfnamefont {A.~C.}\ \bibnamefont {Potter}}, \bibinfo {author} {\bibfnamefont {S.}~\bibnamefont {Gopalakrishnan}}, \ and\ \bibinfo {author} {\bibfnamefont {R.}~\bibnamefont {Vasseur}},\ }\href {\doibase 10.1103/PhysRevLett.129.200602} {\bibfield  {journal} {\bibinfo  {journal} {Phys. Rev. Lett.}\ }\textbf {\bibinfo {volume} {129}},\ \bibinfo {pages} {200602} (\bibinfo {year} {2022}{\natexlab{a}})}\BibitemShut {NoStop}%
\bibitem [{\citenamefont {Ippoliti}\ and\ \citenamefont {Khemani}(2024)}]{PRXQuantum.5.020304}%
  \BibitemOpen
  \bibfield  {author} {\bibinfo {author} {\bibfnamefont {M.}~\bibnamefont {Ippoliti}}\ and\ \bibinfo {author} {\bibfnamefont {V.}~\bibnamefont {Khemani}},\ }\href {\doibase 10.1103/PRXQuantum.5.020304} {\bibfield  {journal} {\bibinfo  {journal} {PRX Quantum}\ }\textbf {\bibinfo {volume} {5}},\ \bibinfo {pages} {020304} (\bibinfo {year} {2024})}\BibitemShut {NoStop}%
\bibitem [{\citenamefont {Agrawal}\ \emph {et~al.}(2023)\citenamefont {Agrawal}, \citenamefont {Lopez-Piqueres}, \citenamefont {Vasseur}, \citenamefont {Gopalakrishnan},\ and\ \citenamefont {Potter}}]{agrawal2023observingquantummeasurementcollapse}%
  \BibitemOpen
  \bibfield  {author} {\bibinfo {author} {\bibfnamefont {U.}~\bibnamefont {Agrawal}}, \bibinfo {author} {\bibfnamefont {J.}~\bibnamefont {Lopez-Piqueres}}, \bibinfo {author} {\bibfnamefont {R.}~\bibnamefont {Vasseur}}, \bibinfo {author} {\bibfnamefont {S.}~\bibnamefont {Gopalakrishnan}}, \ and\ \bibinfo {author} {\bibfnamefont {A.~C.}\ \bibnamefont {Potter}},\ }\href {https://arxiv.org/abs/2311.00058} {\enquote {\bibinfo {title} {Observing quantum measurement collapse as a learnability phase transition},}\ } (\bibinfo {year} {2023}),\ \Eprint {http://arxiv.org/abs/2311.00058} {arXiv:2311.00058 [quant-ph]} \BibitemShut {NoStop}%
\bibitem [{\citenamefont {Noh}\ \emph {et~al.}(2020)\citenamefont {Noh}, \citenamefont {Jiang},\ and\ \citenamefont {Fefferman}}]{Noh2020efficientclassical}%
  \BibitemOpen
  \bibfield  {author} {\bibinfo {author} {\bibfnamefont {K.}~\bibnamefont {Noh}}, \bibinfo {author} {\bibfnamefont {L.}~\bibnamefont {Jiang}}, \ and\ \bibinfo {author} {\bibfnamefont {B.}~\bibnamefont {Fefferman}},\ }\href {\doibase 10.22331/q-2020-09-11-318} {\bibfield  {journal} {\bibinfo  {journal} {{Quantum}}\ }\textbf {\bibinfo {volume} {4}},\ \bibinfo {pages} {318} (\bibinfo {year} {2020})}\BibitemShut {NoStop}%
\bibitem [{\citenamefont {Schuster}\ \emph {et~al.}(2024)\citenamefont {Schuster}, \citenamefont {Yin}, \citenamefont {Gao},\ and\ \citenamefont {Yao}}]{schuster2024polynomialtimeclassicalalgorithmnoisy}%
  \BibitemOpen
  \bibfield  {author} {\bibinfo {author} {\bibfnamefont {T.}~\bibnamefont {Schuster}}, \bibinfo {author} {\bibfnamefont {C.}~\bibnamefont {Yin}}, \bibinfo {author} {\bibfnamefont {X.}~\bibnamefont {Gao}}, \ and\ \bibinfo {author} {\bibfnamefont {N.~Y.}\ \bibnamefont {Yao}},\ }\href {https://arxiv.org/abs/2407.12768} {\enquote {\bibinfo {title} {A polynomial-time classical algorithm for noisy quantum circuits},}\ } (\bibinfo {year} {2024}),\ \Eprint {http://arxiv.org/abs/2407.12768} {arXiv:2407.12768 [quant-ph]} \BibitemShut {NoStop}%
\bibitem [{\citenamefont {Schollw{\"o}ck}(2011)}]{SCHOLLWOCK201196}%
  \BibitemOpen
  \bibfield  {author} {\bibinfo {author} {\bibfnamefont {U.}~\bibnamefont {Schollw{\"o}ck}},\ }\href {\doibase https://doi.org/10.1016/j.aop.2010.09.012} {\bibfield  {journal} {\bibinfo  {journal} {Annals of Physics}\ }\textbf {\bibinfo {volume} {326}},\ \bibinfo {pages} {96} (\bibinfo {year} {2011})},\ \bibinfo {note} {january 2011 Special Issue}\BibitemShut {NoStop}%
\bibitem [{\citenamefont {Sala}\ \emph {et~al.}(2024)\citenamefont {Sala}, \citenamefont {Gopalakrishnan}, \citenamefont {Oshikawa},\ and\ \citenamefont {You}}]{You24weaksym}%
  \BibitemOpen
  \bibfield  {author} {\bibinfo {author} {\bibfnamefont {P.}~\bibnamefont {Sala}}, \bibinfo {author} {\bibfnamefont {S.}~\bibnamefont {Gopalakrishnan}}, \bibinfo {author} {\bibfnamefont {M.}~\bibnamefont {Oshikawa}}, \ and\ \bibinfo {author} {\bibfnamefont {Y.}~\bibnamefont {You}},\ }\href {\doibase 10.1103/PhysRevB.110.155150} {\bibfield  {journal} {\bibinfo  {journal} {Phys. Rev. B}\ }\textbf {\bibinfo {volume} {110}},\ \bibinfo {pages} {155150} (\bibinfo {year} {2024})}\BibitemShut {NoStop}%
\bibitem [{\citenamefont {{Lessa}}\ \emph {et~al.}(2024)\citenamefont {{Lessa}}, \citenamefont {{Ma}}, \citenamefont {{Zhang}}, \citenamefont {{Bi}}, \citenamefont {{Cheng}},\ and\ \citenamefont {{Wang}}}]{Wang24strtowksym}%
  \BibitemOpen
  \bibfield  {author} {\bibinfo {author} {\bibfnamefont {L.~A.}\ \bibnamefont {{Lessa}}}, \bibinfo {author} {\bibfnamefont {R.}~\bibnamefont {{Ma}}}, \bibinfo {author} {\bibfnamefont {J.-H.}\ \bibnamefont {{Zhang}}}, \bibinfo {author} {\bibfnamefont {Z.}~\bibnamefont {{Bi}}}, \bibinfo {author} {\bibfnamefont {M.}~\bibnamefont {{Cheng}}}, \ and\ \bibinfo {author} {\bibfnamefont {C.}~\bibnamefont {{Wang}}},\ }\href@noop {} {\bibfield  {journal} {\bibinfo  {journal} {arXiv}\ } (\bibinfo {year} {2024})},\ \Eprint {http://arxiv.org/abs/2405.03639} {2405.03639} \BibitemShut {NoStop}%
\bibitem [{\citenamefont {Aaronson}(2005)}]{doi:10.1098/rspa.2005.1546}%
  \BibitemOpen
  \bibfield  {author} {\bibinfo {author} {\bibfnamefont {S.}~\bibnamefont {Aaronson}},\ }\href {\doibase 10.1098/rspa.2005.1546} {\bibfield  {journal} {\bibinfo  {journal} {Proceedings of the Royal Society A: Mathematical, Physical and Engineering Sciences}\ }\textbf {\bibinfo {volume} {461}},\ \bibinfo {pages} {3473} (\bibinfo {year} {2005})},\ \Eprint {http://arxiv.org/abs/https://royalsocietypublishing.org/doi/pdf/10.1098/rspa.2005.1546} {https://royalsocietypublishing.org/doi/pdf/10.1098/rspa.2005.1546} \BibitemShut {NoStop}%
\bibitem [{\citenamefont {Fishman}\ \emph {et~al.}(2022{\natexlab{a}})\citenamefont {Fishman}, \citenamefont {White},\ and\ \citenamefont {Stoudenmire}}]{ITensor}%
  \BibitemOpen
  \bibfield  {author} {\bibinfo {author} {\bibfnamefont {M.}~\bibnamefont {Fishman}}, \bibinfo {author} {\bibfnamefont {S.~R.}\ \bibnamefont {White}}, \ and\ \bibinfo {author} {\bibfnamefont {E.~M.}\ \bibnamefont {Stoudenmire}},\ }\href {\doibase 10.21468/SciPostPhysCodeb.4} {\bibfield  {journal} {\bibinfo  {journal} {SciPost Phys. Codebases}\ ,\ \bibinfo {pages} {4}} (\bibinfo {year} {2022}{\natexlab{a}})}\BibitemShut {NoStop}%
\bibitem [{\citenamefont {Fishman}\ \emph {et~al.}(2022{\natexlab{b}})\citenamefont {Fishman}, \citenamefont {White},\ and\ \citenamefont {Stoudenmire}}]{ITensor-r0.3}%
  \BibitemOpen
  \bibfield  {author} {\bibinfo {author} {\bibfnamefont {M.}~\bibnamefont {Fishman}}, \bibinfo {author} {\bibfnamefont {S.~R.}\ \bibnamefont {White}}, \ and\ \bibinfo {author} {\bibfnamefont {E.~M.}\ \bibnamefont {Stoudenmire}},\ }\href {\doibase 10.21468/SciPostPhysCodeb.4-r0.3} {\bibfield  {journal} {\bibinfo  {journal} {SciPost Phys. Codebases}\ ,\ \bibinfo {pages} {4}} (\bibinfo {year} {2022}{\natexlab{b}})}\BibitemShut {NoStop}%
\bibitem [{\citenamefont {Vidal}(2003)}]{PhysRevLett.91.147902}%
  \BibitemOpen
  \bibfield  {author} {\bibinfo {author} {\bibfnamefont {G.}~\bibnamefont {Vidal}},\ }\href {\doibase 10.1103/PhysRevLett.91.147902} {\bibfield  {journal} {\bibinfo  {journal} {Phys. Rev. Lett.}\ }\textbf {\bibinfo {volume} {91}},\ \bibinfo {pages} {147902} (\bibinfo {year} {2003})}\BibitemShut {NoStop}%
\bibitem [{\citenamefont {Barratt}\ \emph {et~al.}(2022{\natexlab{b}})\citenamefont {Barratt}, \citenamefont {Agrawal}, \citenamefont {Gopalakrishnan}, \citenamefont {Huse}, \citenamefont {Vasseur},\ and\ \citenamefont {Potter}}]{PhysRevLett.129.120604}%
  \BibitemOpen
  \bibfield  {author} {\bibinfo {author} {\bibfnamefont {F.}~\bibnamefont {Barratt}}, \bibinfo {author} {\bibfnamefont {U.}~\bibnamefont {Agrawal}}, \bibinfo {author} {\bibfnamefont {S.}~\bibnamefont {Gopalakrishnan}}, \bibinfo {author} {\bibfnamefont {D.~A.}\ \bibnamefont {Huse}}, \bibinfo {author} {\bibfnamefont {R.}~\bibnamefont {Vasseur}}, \ and\ \bibinfo {author} {\bibfnamefont {A.~C.}\ \bibnamefont {Potter}},\ }\href {\doibase 10.1103/PhysRevLett.129.120604} {\bibfield  {journal} {\bibinfo  {journal} {Phys. Rev. Lett.}\ }\textbf {\bibinfo {volume} {129}},\ \bibinfo {pages} {120604} (\bibinfo {year} {2022}{\natexlab{b}})}\BibitemShut {NoStop}%
\end{thebibliography}%




\end{document}